\documentclass[aps,prd,10pt,showpacs,amsmath,twocolumn,floatfix,amssymb,nofootinbib,longbibliography]{revtex4-2}

\usepackage{graphicx}
\usepackage{comment}
\usepackage[usenames]{color}
\usepackage{bm}
\usepackage{ifpdf}
\usepackage{floatrow}
\usepackage{makecell}
\usepackage{caption}
\usepackage{stackrel}
\usepackage{subcaption}
\usepackage{enumitem}
\usepackage{babel}

\usepackage[normalem]{ulem}
\usepackage[dvipsnames]{xcolor}
\usepackage[utf8]{inputenc}
\usepackage{hyperref}
\hypersetup{
  pdfnewwindow=true,      
  colorlinks=true,        
  linkcolor=PineGreen,    
  citecolor=PineGreen,    
  filecolor=PineGreen,    
  urlcolor=PineGreen      
}
\newcommand{\be}{\begin{eqnarray}}
\newcommand{\ee}{\end{eqnarray}}

\usepackage{parskip}
\usepackage{anyfontsize}
\begin{document}

\title{Toward extracting scattering phase shift from integrated correlation functions IV: Coulomb corrections}

\author{Peng~Guo}
\email{peng.guo@dsu.edu}

\affiliation{College of Arts and Sciences,  Dakota State University, Madison, SD 57042, USA}
\affiliation{Kavli Institute for Theoretical Physics, University of California, Santa Barbara, CA 93106, USA}
 
\author{Frank~X.~Lee}
\email{fxlee@gwu.edu}
\affiliation{Department of Physics, George Washington University, Washington, DC 20052, USA}

\date{\today}

\begin{abstract}
The  formalism developed in Refs.~\cite{Guo:2023ecc,Guo:2024zal,Guo:2024pvt}  that  relates  the integrated correlation functions for a trapped system to the infinite volume scattering phase shifts through a weighted integral is further extended to include Coulomb interaction between charged particles. The original formalism cannot be applied due to different divergent asymptotic behavior resulting from  the long-range nature of the Coulomb force.  We show that a modified formula in which the difference of  integrated correlation functions between particles interacting with Coulomb plus short-range interaction and with Coulomb interaction alone is free of divergence, and has rapid approach to its infinite volume limit. Using an exactly solvable model, we demonstrate that the short-range potential scattering phase shifts can be reliably extracted from the formula in the presence of Coulomb interaction. 
\end{abstract}

\maketitle

\section{Introduction}\label{sec:intro}

Scattering plays a crucial role in a wide range of dynamics, from the strong interaction in quantum chromodynamics  (QCD)    to atomic interactions in condensed matter physics.  Precisie determination of scattering amplitudes in such systems remains fundamental but challenging. In most cases, numerical simulations based on stochastic evaluation of the path integral are performed by placing the system in artificial traps, such as a periodic finite box or a harmonic oscillator trap. The traps lead to quantized energies in the system, which are then  connected  to the infinite volume scattering amplitudes through quantization conditions, such as the  L\"uscher formula  \cite{Luscher:1990ux} for periodic boxes, and  Busch-Englert-Rza\.zewski-Wilkens (BERW) formula \cite{Busch98}  in harmonic oscillator traps. A lot of progress has been made in recent years on  extracting multi-hadron dynamics in nuclear physics, where  L\"uscher-  or BERW-like formula  has been successfully extended into inelastic channel, three-body channel, and other systems, see e.g. Refs.~\cite{Rummukainen:1995vs,Christ:2005gi,Bernard:2008ax,He:2005ey,Lage:2009zv,Doring:2011vk,Guo:2012hv,Guo:2013vsa,Kreuzer:2008bi,Polejaeva:2012ut,Hansen:2014eka,Mai:2017bge,Mai:2018djl,Doring:2018xxx,Guo:2016fgl,Guo:2017ism,Guo:2017crd,Guo:2018xbv,Mai:2019fba,Guo:2018ibd,Guo:2019hih,Guo:2019ogp,Guo:2020wbl,Guo:2020kph,Guo:2020iep,Guo:2020ikh,Guo:2020spn,Guo:2021lhz,Guo:2021uig,Guo:2021qfu,Guo:2021hrf,Stetcu:2007ms, Stetcu:2010xq,Rotureau:2010uz,Rotureau:2011vf,Luu:2010hw,Yang:2016brl,Johnson:2019sps,Zhang:2019cai, Zhang:2020rhz}.
In addition to the L\"{u}scher-like method, there is the HALQCD potential method  \cite{PhysRevLett.99.022001,10.1143/PTP.123.89,PhysRevD.99.014514,ISHII2012437,AokiEPJA2013}  that also relies on the discrete energy spectrum.  

In multi-nucleon systems, issues such as the signal-to-noise (S/N)  ratio  of lattice correlation functions \cite{lepage1989analysis,DRISCHLER2021103888}  and the requirement of increasingly large number of interpolating operators  at large volumes \cite{Bulava:2019kbi}, present unique challenges to L\"{u}scher-like methods. These challenges motivated alternative approaches.    To this end, the integrated correlation function method  was proposed recently  in Ref.~\cite{Guo:2023ecc}. It relates  the difference between interacting and non-interacting integrated correlation functions  of two non-relativistic particles in an artificial trap  to the  infinite-volume phase shift through a weighted integral. The main advantage of the method is working directly with correlation functions, bypassing the energy spectrum determination. Furthermore, the relation  has a rapid  convergence rate at short Euclidean times, even with a modestly small sized trap~\cite{Guo:2023ecc}. This makes it  potentially  a good candidate to overcome the S/N problem in multi-nucleon systems.  The formalism is by construction free from issues encountered at large volumes, such as increasingly dense energy spectrum and the extraction of low-lying states. The integrated correlation function formalism was later extended to include relativistic dynamics \cite{Guo:2024zal},  coupled channel effects in \cite{Guo:2024bar,Guo:2024pvt}, and its potential simulation on quantum computers \cite{Guo:2025vgk}.

 The aim of the present work is to further develop the integrated correlation function formalism to include long-range Coulomb interaction between charged particles, in conjunction with short-range interactions. It is a necessary step to describe charged-hadron interactions in nuclear physics and lattice QCD.  Coulomb corrections have been considered in traditional formalisms in harmonic traps~\cite{Zhang:2024mot,Bagnarol:2024rhq,Zhang:2024vch}, 
 and periodic boxes ~\cite{Beane_2014,NPLQCD:2020ozd,Yu:2022nzm,bubna2024}.

 In the integrated correlation function formalism considered here,
  due to the distortion of long-range Coulomb interaction on the asymptotic wavefunction, the Coulomb-modified  integrated correlation function has a different divergent asymptotic behavior from that of the non-interacting integrated correlation function. Consequently, the difference between the Coulomb-modified  integrated correlation function and the non-interacting integrated correlation function in the original formalism is cutoff dependent and divergent, rendering it inapplicable to scattering problems with Coulomb interaction.  We will show that the issue can be resolved by carefully incorporating the asymptotic behavior of the correlation function due to pure Coulomb interaction.

The paper is organized as follows. A brief summary of integrated correlation function formalism is outlined in Sec.~\ref{sec:sumcorrformula}.  The extension to Coulomb interaction  is presented in  Sec.~\ref{sec:coulombcorrformula}.  A numerical test with an exactly solvable contact interaction in a spherical hard-wall trap is discussed in  Sec.~\ref{sec:example}, followed by summary and outlook in Sec.~\ref{sec:summary}. Some technical details are in two appendices.

\section{Integrated correlation function formalism without Coulomb interaction}\label{sec:sumcorrformula}  
 In this section, we outline the essential ingredients of  the  integrated correlation function formalism needed in the discussion of Coulomb interaction in the next section.
 
A relation in $1+1$ dimensional spacetime that connects the  integrated    correlation functions for  two non-relativistic  particles in a trap to the scattering phase shift due to a short-range interaction potential, $\delta (\epsilon)$, through a weighted integral,  was derived in  Ref.~\cite{Guo:2023ecc},
  \begin{equation}
 C(t) - C^{(0)} (t) \stackrel{\text{trap} \rightarrow \infty}{\rightarrow} \frac{1}{\pi} \int_0^{\infty} d \epsilon \, \frac{d  \delta(\epsilon) }{d \epsilon}\, e^{- i \epsilon t} + \frac{\delta(0)}{\pi},
 \label{single2}
 \end{equation} 
 where $C(t)$ and $C^{(0)} (t)$ are Minkowski time integrated correlation functions for two interacting and non-interacting particles in the trap.

The $3+1$ dimensional spacetime extension of Eq.(\ref{single2}) is given in Eq.(\ref{Ctdiff})  by, see derivations below,  
  \begin{equation}
  C_l(t) - C^{(0)}_l (t) \stackrel{\text{trap} \rightarrow \infty}{\rightarrow} (2l+1) \frac{ i \, t}{\pi} \int_0^{\infty} d \epsilon  \,  \delta^{(S)}_l(\epsilon)\,  e^{- i \epsilon t}  ,
 \label{Ctpartialwav}
 \end{equation} 
where $C_l(t)$ and $C_l^{(0)} (t)$ are partial-wave-projected integrated correlation functions of angular momentum-$l$, and $\delta^{(S)}_l (\epsilon)$ stands for the $l$-th partial-wave scattering phase shift due to a short-range interaction potential (emphasized by the superscript-$S$). We also assume that $\delta^{(S)}_l (0) =0$. The extra factor $(2l+1)$ is the result of partial-wave projection. We remark that  at current scope of presentation, we will limit ourself to the traps that preserve rotational symmetry, such as harmonic oscillator trap and spherical hard-wall trap,  so that angular momenta are  good   quantum numbers. The partial-wave expansion of trapped wavefunction and Green's function depends only on orbital quantum number-$l$, not magnetic quantum number-$m_l$ where $m_l \in [ - l, \cdots, l ]$. In cases where continuous rotational symmetry is no longer the symmetry group of the trap, such as a periodic cubic box, the dynamic equations of a trapped system  has to be projected into the  irreducible representations (irreps) of the cubic symmetry group. The projection of each irrep of the cubic symmetry group typically involves the mixture of  angular momentum partial waves, see e.g. \cite{Doring:2018xxx,Cornwell:1997ke}.  Typically the higher angular momentum components could be neglected due to the threshold barrier factor suppression at low energy and exponential factor in the integrand of the Euclidean time, and irrep projection is dominated by the few lowest angular momentum components. Similar effects are expected in the integrated correlation function approach. On top of that, due to the suppression of the extra relativistic kinematic factor and exponential decaying factor in the Euclidean time, see e.g. Ref.~\cite{Guo:2024zal},  the inifinite volume limit of the difference of integrated correlation functions should still be dominated by low-energy dynamics.   We leave the technical aspects of irreps projection for periodic box traps for future discussion. 

 The integrated forward time propagating two-particle correlation function for non-relativistic systems is defined through summing over all the modes of  two-particle correlation functions  along the diagonal,
\begin{equation}
C_l(t) = (2 l+1) \int_0^\infty r^2 d  r   \langle 0 |  \widehat{\mathcal{O}}_{l } (r, t) \widehat{\mathcal{O}}_{l }^\dag (r, 0)   | 0 \rangle. \label{Ctdef}
\end{equation}
The  $\widehat{\mathcal{O}}_l^\dag (r, 0)$ and $\widehat{\mathcal{O}}_l (r, t)$ denote creation  and annihilation operators to create two particles  with relative radial coordinate $r$ and angular momentum-$l$ at time $0$ at the source, and then to annihilate them with relative radial coordinate of $r$ at later time $t$ at the sink, respectively.   Examples of construction of two-particle creation operators in $1+1$ dimensions can be found in Eq.(5) in  Ref.~\cite{Guo:2024pvt}, or Eq.(4) in Ref.~\cite{Guo:2025vgk}.

 We also showed in Ref.~\cite{Guo:2023ecc} that   two-particle correlation functions can be expressed in terms of wavefunctions in the spectral representation.  By inserting a complete energy basis in between two-particle creation and annihilation operators: $\sum_\epsilon | \epsilon \rangle \langle \epsilon | =1$, the wavefunction defined via,  
\begin{equation}
 \langle \epsilon |  \widehat{\mathcal{O}}_l^\dag (r, 0) | 0 \rangle  = \psi_l^* (r; \epsilon) ,
\end{equation} 
 satisfies the radial Schr\"odinger equation for a trapped system (we use a dimensionless unit system in which $\hbar=c=1$),
\begin{align}
& \left  [  -  \frac{1}{2\mu r^2} \frac{d}{d r} \left ( r^2 \frac{d }{d r}  \right ) + \frac{l (l+1)}{ 2\mu r^2} +  U_{trap} (r) +V (r)    \right ] \psi_l(r; \epsilon)  \nonumber \\
& =\epsilon \psi_l(r; \epsilon), \label{schrodingereqtrap}
\end{align}
where $\mu$ denotes the reduced mass of the two-particle system,  and $U_{trap} (r)$ and $V (r)=V_S(r)$  are  trap potential and  short-range two-particle interaction potential, respectively. Thus in spectral representation the  integrated non-relativistic two-particle correlation function is given by~\cite{Guo:2023ecc}, 
\begin{equation}
C_l(t)  =  (2 l+1) \int_0^\infty r^2 d  r   \sum_\epsilon \psi_l (r; \epsilon) \psi^*_l (r; \epsilon)  e^{- i \epsilon t}     .
\label{Crr}
\end{equation}

 A  relatively simple way to show the connection between Eq.(\ref{Crr}) and Eq.(\ref{Ctpartialwav}) is to consider the asymptotic behavior of infinite-volume wavefunctions,
 \begin{equation}
  \psi^{(\infty)}_l (r; \epsilon)  \stackrel{r\rightarrow \infty}{\rightarrow} \frac{4\pi i^l}{k r} \sin \left ( k r - \frac{\pi l}{2} + \delta^{(S)}_l (\epsilon) \right ), \label{partialwavasymp}
 \end{equation}
 where $k = \sqrt{2\mu \epsilon}$ is the relative momentum in the center of mass. 
 Using the identity relation (see Eq.(31) of Ref.~\cite{Poliatzky:1992gn}) involving the radial wavefunction,
 \begin{align}
2k  \int_0^\Lambda d r  | u^{(\infty)}_l (r; k)  |^2 &=   \partial_\Lambda u^{(\infty)*}_l (\Lambda ; k)  \partial_k u^{(\infty)}_l (\Lambda ; k)   \nonumber \\
&  -  u^{(\infty) *}_l (\Lambda ; k)   \partial_k \partial_\Lambda u^{(\infty)}_l (\Lambda ; k)  , \label{uidentity}
 \end{align}
 where $\Lambda \rightarrow \infty$  is a distance that is far larger than the potential range, 
 and its relation to the asymptotic form in Eq.(\ref{partialwavasymp}), 
 \begin{equation}
 u^{(\infty)}_l (r; k)  = \frac{k r}{4\pi i^l}  \psi^{(\infty)}_l (r; \epsilon) ,
 \end{equation}
 we find that
 \begin{equation}
  \int_0^\Lambda  d  r   |  u^{(\infty)}_l (r; k)  |^2   \stackrel{\Lambda \rightarrow \infty}{\rightarrow}   \frac{\Lambda}{2} + \frac{\sin \left ( 2 k \Lambda - \pi l   \right )}{4 k}  + \frac{1}{2} \frac{d \delta^{(S)}_l (\epsilon) }{d k} .
\label{integratedwav}
\end{equation}
Its corresponding non-interacting form is given by, 
 \begin{equation}
  \int_0^\Lambda  d  r   |  u^{(0, \infty)}_l (r; k)  |^2   \stackrel{\Lambda \rightarrow \infty}{\rightarrow}   \frac{\Lambda}{2} + \frac{\sin \left ( 2 k \Lambda - \pi l   \right )}{4 k}    .
\label{integratedwav0}
\end{equation}
We see that Eq.(\ref{integratedwav}) and Eq.(\ref{integratedwav0}) have the same divergent asymptotic terms, and they cancel out in the difference,
 \begin{equation}
  \int_0^\Lambda  d  r  \left [   |  u^{(\infty)}_l (r; k)  |^2 -|  u^{(0, \infty)}_l (r; k)  |^2   \right ]   \stackrel{\Lambda \rightarrow \infty}{\rightarrow}    \frac{1}{2} \frac{d \delta^{(S)}_l (\epsilon) }{d k} ,
\label{diffintegratedwav}
\end{equation}
which remains finite and is free of the integration cutoff $\Lambda$.
Therefore, at infinite volume limit,  the difference of integrated correlation functions between interacting and non-interacting cases takes the form, 
   \begin{align}
 & C_l(t) - C^{(0)}_l (t) 
  \stackrel{\text{trap} \rightarrow \infty}{\rightarrow} (2l+1)  \int_0^\infty \frac{k^2 d k}{(2\pi)^3} \nonumber \\
 & \times  \frac{(4\pi)^2}{k^2}   \int_0^\Lambda  d  r  \left [   |  u^{(\infty)}_l (r; k)  |^2 -|  u^{(0, \infty)}_l (r; k)  |^2   \right ]   e^{- i \frac{k^2}{2\mu} t}  \nonumber \\
& \stackrel{\Lambda \rightarrow \infty}{\rightarrow}  \frac{(2l+1)}{\pi} \int_0^{\infty} d \epsilon  \frac{d  \delta^{(S)}_l(\epsilon) }{d \epsilon} e^{- i \epsilon t} ,
\label{Ctdiff}
 \end{align} 
which yields Eq.(\ref{Ctpartialwav}) after integration by part.

A more rigorous way of proving Eq.(\ref{Ctpartialwav}) is to use the Green's function representation of integrated correlation functions, see technical details in Ref.~\cite{Guo:2023ecc,Guo:2024pvt},
\begin{equation}
C_l(t) = i \int_{-\infty}^\infty \frac{d \lambda}{2\pi} Tr \left [ G_l^{(trap)} ( \lambda  ) \right ] e^{- i \lambda t}, \label{Ctgreen}
\end{equation} 
  The partial-wave Green's function  $G_l^{(trap)}$ is the solution of partial-wave Dyson equation,  whose spectral representation is given by, 
\begin{equation}
G_l^{(trap)} ( r, r' ; \lambda ) = \sum_{\epsilon} \frac{\psi_l (r; \epsilon) \psi^*_l (r';\epsilon) }{  \lambda - \epsilon } .
\end{equation} 
The trace   is defined by,
\begin{equation}
 Tr \left [ G_l^{(trap)} ( \lambda  ) \right ]  = \int_0^\infty r^2 d r G_l^{(trap)} (r,r ; \lambda  ).
\end{equation} 
At infinite volume limit,  the difference between trace of Green's functions  of interacting and non-interacting systems is related to   scattering phase shift through a dispersion integral, see e.g. Refs.~\cite{Guo:2022row,Guo:2023ecc,Guo:2024bar},
\begin{equation} 
 Tr  \left [ G_l^{(\infty)} (  \epsilon) -G_l^{(0, \infty)} (    \epsilon )   \right ]  = - \frac{1}{\pi}  \int_{0}^{\infty} d \lambda  \frac{\delta^{(S)}_l  (\lambda)}{ (\lambda - \epsilon)^2}. \label{Friedelformula}
\end{equation}
This relation is the result of Friedel formula \cite{Friedel1958} and Krein’s theorem \cite{zbMATH03313022,krein1953trace}. 
Eq.(\ref{Ctgreen}) and Eq.(\ref{Friedelformula})  together lead to the same relation in Eq.(\ref{Ctpartialwav}), also see discussions in Ref.~\cite{Guo:2023ecc,Guo:2024pvt}.

\section{Integrated correlation function formalism with Coulomb interaction}\label{sec:coulombcorrformula}  
When the long-range Coulomb interaction is involved, the original relation in Eq.(\ref{Ctpartialwav}) cannot be applied directly, mainly due to the distortion in  the asymptotic wavefunction, see e.g. Ref.~\cite{messiah1999quantum}. 
In this section, we explain what issues may arise from the inclusion of  Coulomb interaction and how we can overcome the issues and  modify Eq.(\ref{Ctpartialwav}).

The total interaction potential has two contributions,   
\begin{equation}
V(r) = V_S (r) +V_C (r), 
\end{equation}
where $V_S (r)$ stands for a short-range potential, and $V_C (r)$  a long-range Coulomb potential between two electric charges $Z_1$ and $Z_2$,
\begin{equation}
V_C(r) = -\frac{Z}{r}, \text{ with } Z= -Z_1 Z_2 e^2,
\end{equation}
whose strength is represented by $Z$. The asymptotic behavior of infinite volume wavefunction is now given by, see e.g. Ref.~\cite{Guo:2021qfu},
 \begin{equation}
  \psi^{(\infty)}_l (r; \epsilon)  \stackrel{r\rightarrow \infty}{\rightarrow} \frac{4\pi i^l}{k r} \sin \left ( k r - \frac{\pi l}{2} + \delta_l (\epsilon) - \gamma \ln (2 k r) \right ), \label{SCwavasymp}
 \end{equation}
where $\gamma =- \frac{Z\mu}{k} $.   The total partial phase shift, 
\begin{equation}
 \delta_l (\epsilon)  =  \delta^{(S)}_l (\epsilon)  +  \delta^{(C)}_l (\epsilon) ,
\end{equation}
is the sum of short-range and long-range phase shifts. The Coulomb phase shift has the analytic expression in terms of gamma function $\Gamma$, see e.g. Ref.~\cite{Guo:2021qfu}, 
\begin{equation}
\delta^{(C)}_l (\epsilon) = \arg \Gamma (l +1  + i \gamma ) . 
\label{Coulombphaseshift}
\end{equation}
Using the distorted infinite volume asymptotic wavefunction in Eq.(\ref{SCwavasymp}), and the identity relation in Eq.(\ref{uidentity}),  we find
 \begin{align}
 &  \int_0^\Lambda  d  r   |  u^{(\infty)}_l (r; k)  |^2   \stackrel{\Lambda \rightarrow \infty}{\rightarrow}   \frac{\Lambda - \frac{d}{d k} \left (  \gamma \ln (2 k \Lambda)  \right )}{2}  \nonumber \\
 & + \frac{\sin \left ( 2 k \Lambda - \pi l  - 2 \gamma \ln (2 k \Lambda)  \right )}{4 k}  + \frac{1}{2} \frac{d \delta_l (\epsilon) }{d k} .
\label{integratedwavSC}
\end{align}
Clearly Eq.(\ref{integratedwavSC}) has a different divergent asymptotic behavior from Eq.(\ref{integratedwav0}). 
Their difference is cutoff $\Lambda$ dependent and diverges as $\Lambda \rightarrow \infty$. This is the primary reason that  Eq.(\ref{Ctpartialwav}) cannot be applied  directly when long-range Coulomb interaction is involved.

Fortunately, the asymptotic wavefunction with Coulomb interaction potential alone~\cite{messiah1999quantum,Guo:2021qfu},
 \begin{align}
&  \psi^{(C, \infty)}_l (r; \epsilon) \nonumber \\
&  \stackrel{r\rightarrow \infty}{\rightarrow} \frac{4\pi i^l}{k r} \sin \left ( k r - \frac{\pi l}{2} + \delta^{(C)}_l (\epsilon) - \gamma \ln (2 k r) \right ), \label{Coulombwavasymp}
 \end{align}
 has the exact same form as Eq.(\ref{SCwavasymp}) except with $\delta_l (\epsilon) $ replaced by $\delta^{(C)}_l (\epsilon)$.
As a result, with only Coulomb interaction, we find 
 \begin{align}
 &  \int_0^\Lambda  d  r   |  u^{(C,\infty)}_l (r; k)  |^2   \stackrel{\Lambda \rightarrow \infty}{\rightarrow}   \frac{\Lambda - \frac{d}{d k} \left (  \gamma \ln (2 k \Lambda)  \right )}{2}  \nonumber \\
 & + \frac{\sin \left ( 2 k \Lambda - \pi l  - 2 \gamma \ln (2 k \Lambda)  \right )}{4 k}  + \frac{1}{2} \frac{d \delta^{(C)}_l (\epsilon) }{d k} .
\label{integratedwavCoulomb}
\end{align}
Therefore, the difference between Eq.(\ref{integratedwavSC})  and Eq.(\ref{integratedwavCoulomb}) yields the finite result,
 \begin{equation}
  \int_0^\Lambda  d  r  \left [   |  u^{(\infty)}_l (r; k)  |^2 -|  u^{(C, \infty)}_l (r; k)  |^2   \right ]   \stackrel{\Lambda \rightarrow \infty}{\rightarrow}    \frac{1}{2} \frac{d \delta^{(S)}_l (\epsilon) }{d k}.
\label{diffintegratedwavSC}
\end{equation}
Using the same steps as in Eq.\eqref{diffintegratedwav} and Eq.\eqref{Ctdiff} and integration by parts,  we arrive at the final expression  in {\em Minkowski spacetime}, 
  \begin{equation}
 C_l(t) - C^{(C)}_l (t) \stackrel{\text{trap} \rightarrow \infty}{\rightarrow} (2l+1) \frac{ i \, t}{\pi} \int_0^{\infty} d \epsilon  \,  \delta^{(S)}_l(\epsilon)\,  e^{- i \epsilon t}  ,
 \label{CtpartialwavSC}
 \end{equation} 
Its counterpart in  {\em Euclidean spacetime} can be obtained by an analytic continuation $t \to - i \tau $,
  \begin{equation}
    C_l(\tau) - C^{(C)}_l (\tau) \stackrel{\text{trap} \rightarrow \infty}{\rightarrow} (2l+1) \frac{ \tau}{\pi} \int_0^{\infty} d \epsilon  \,  \delta^{(S)}_l(\epsilon)\, e^{- \epsilon \tau}.
 \label{CtpartialwavSCE}
 \end{equation} 
Eq.(\ref{CtpartialwavSC}) bears a close resemblance to  Eq.(\ref{Ctpartialwav}) in the absence of Coulomb interaction, except that the left-hand side now takes on new meanings. 
 The $C_l (t)$ is the integrated correlation function for trapped particles interacting with both short-range potential and long-range Coulomb interaction, $C^{(C)}_l (t)$ with only Coulomb interaction instead of non-interacting $C^{(0)}_l (t)$, while $\delta^{(S)}_l(\epsilon)$ is the infinite volume phase shift from the  short-range interaction only.   Once $\delta^{(S)}_l$  is extracted from Eq.(\ref{CtpartialwavSC}), the total phase shift is simply given by adding to it  the Coulomb phase shift in  Eq.(\ref{Coulombphaseshift}).

\section{Numerical Verification  with an exactly solvable model }\label{sec:example}  
Having derived the Coulomb-modified relation in Eq.\eqref{CtpartialwavSC} or Eq.\eqref{CtpartialwavSCE}, it is important to check its validity. To this end, we employ an exactly solvable model in which every  aspect of the problem is known  in closed form.

For the short-range interaction, we adopt a contact interaction potential,
\begin{equation}
V_S(\mathbf{ r}) = V_0  \frac{ \delta(r) }{r^2},
\end{equation}
where $V_0$ is the bare potential strength.
With a contact interaction, only $S$-wave ($l=0$) phase shift  will contribute, and its analytical solution in 3D is outlined in Appendix \ref{scatteringsolutions}.  
The phase shift is given by,
\begin{equation}
\delta^{(S)}_0 (\epsilon) = \cot^{-1} \left ( - \frac{1}{2\mu k V_R} \right ) \stackrel{V_R \rightarrow 0}{\rightarrow} 0,
\label{phase}
\end{equation}
where $V_R$ denotes renormalized contact interaction potential strength in Eq.\eqref{VR}. 
The integration on the right-hand side of Eq.(\ref{CtpartialwavSC}) can be carried out,
  \begin{equation}
   \frac{ i  t}{\pi} \int_0^{\infty} d \epsilon    \delta^{(S)}_0 (\epsilon)  e^{- i \epsilon t}  = - \frac{1}{2} \mbox{erfc} \left ( \frac{1}{2\mu V_R} \sqrt{\frac{i t}{2\mu}} \right ) e^{ \frac{1}{(2\mu V_R)^2}  \frac{ i t}{2\mu}}.
 \end{equation} 
 in terms of complementary error function $\text{erfc}(z)=1-\text{erf}(z)$.

For the left-hand side of Eq.(\ref{CtpartialwavSC}), 
we need to specify the trap and boundary conditions for the system which will lead to quantized energies. 
The difference of integrated correlation functions can be computed from two energy spectra by,
\begin{equation}
C_l(t) -  C_l^{(C)} (t) = \sum_n \left [ e^{- i \epsilon_n t} - e^{- i \epsilon^{(C)}_n t}  \right ].
\end{equation}
The discrete energy levels $\epsilon_n$ and $\epsilon_n^{(C)}$ are the eigen-energies of Schr\"{o}dinger equation in  Eq.(\ref{schrodingereqtrap}) for a trapped system with short-range plus Coulomb potential and   pure Coulomb potential, respectively.  Alternatively, the discrete energy levels can be obtained  from Coulomb-modified L\"uscher  \cite{Luscher:1990ux}  or  BERW  \cite{Busch98}   formula-like quantization conditions that connect short-range potential phase shifts  with energy levels in the presence of Coulomb force, 
 \begin{equation}
  \det   \left  [  \delta_{lm_l, l'm'_l}        \cot \delta^{(S)}_l (\epsilon)  - \mathcal{M}^{(C)}_{lm_l, l'm'_l}  (\epsilon)    \right  ]   =0 . \label{QCtrap}
\end{equation}  
The  definition of generalized zeta function in the presence of Coulomb force, $\mathcal{M}^{(C)}_{lm_l, l'm'_l}  (\epsilon)  $, can be found in Eq.(59) in Ref.~\cite{Guo:2021qfu}. 
In general, for commonly used traps, such as periodic finite box and harmonic oscillator traps, $\mathcal{M}^{(C)}_{lm_l, l'm'_l}  (\epsilon)  $ has to be solved numerically, which is a highly non-trivial task, see detailed discussion in  Ref.~\cite{Guo:2021qfu}. 

For our purposes, we consider a simple trap: spherical hard-wall with radius $R$,
\begin{equation}
U_{trap} (r) = \begin{cases} 0, &  r <R; \\ \infty, & \text{otherwise}.  \end{cases}.
\end{equation}
In this trap, the quantization condition Eq.(\ref{QCtrap}) is reduced to a simple form,
 \begin{equation}
      \cot \delta^{(S)}_l (\epsilon)   =   \frac{ n_l^{(C)} ( \gamma, k R) }{ j_l^{(C)} ( \gamma, k R) }. \label{qchsw}
\end{equation} 
The derivation is detailed in Appendix \ref{traptoscattappend}, including the 
Coulomb-modified  spherical Bessel functions   $j_l^{(C)}(\gamma, k r)$ and $n_l^{(C)}(\gamma, k r) $  in Eq.(\ref{jLC}) and Eq.(\ref{nLC}).

So the demonstration  boils down to  verifying the following relation,
\begin{align}
& C_0(\tau) -  C_0^{(C)} (\tau) = \sum_n \left [ e^{-  \epsilon_n \tau} - e^{-  \epsilon^{(C)}_n \tau}  \right ] \nonumber \\
& \stackrel{R \rightarrow \infty}{\rightarrow} - \frac{1}{2} \mbox{erfc} \left ( \frac{1}{2\mu V_R} \sqrt{\frac{\tau}{2\mu}} \right ) e^{ \frac{1}{(2\mu V_R)^2}  \frac{ \tau}{2\mu}}. \label{numericeq}
\end{align}
We choose to work in Euclidean spacetime because of better convergence from exponential falloffs as opposed to oscillatory behavior in Minkowski spacetime.
The  $\epsilon_n$ are solved by  Eq.(\ref{qchsw}) with $l=0$, or equivalently
 \begin{equation}
  \delta_0^{(S)} (\epsilon_n ) +\phi_0 (\epsilon_n )  = n \pi, \ \ \ n \in [0,1, \cdots , \infty ] , \label{qcVR}
\end{equation} 
where we introduce a `trapped phase angle' $\phi_0$ to rival the phase shift $ \delta_0^{(S)} $ in infinite volume  for notational convenience,
\begin{equation}
\phi_0 (\epsilon)  \equiv \cot^{-1} \left ( -   \frac{ n_0^{(C)} ( \gamma, k R) }{ j_0^{(C)} ( \gamma, k R) }  \right )   \stackrel{\gamma \rightarrow 0}{\rightarrow}   \cot^{-1} \left ( -  \frac{ n_0  (   k R) }{ j_0  (  k R) }  \right ). 
\label{phi0}
\end{equation}
The   $\epsilon_n^{(C)}$ with pure Coulomb interaction is solved by the quantization condition,
 \begin{equation}
   \phi_0 (\epsilon^{(C)}_n) = n \pi, \ \ \ n \in [0,1, \cdots , \infty ] . \label{qcCoulomb}
\end{equation}  
 The pole position of $\cot \phi_0  $ must be computed numerically by using  Eq.(\ref{qcCoulomb}), however at the limit of $\gamma \rightarrow 0$ (a pure spherical hard wall trap), the pole positions are simply given by $\epsilon_n = \frac{1}{2\mu} (\frac{n \pi}{R})^2$, where $n \in [0, 1, \cdots]$.

  \begin{figure}
\includegraphics[width=0.99\textwidth]{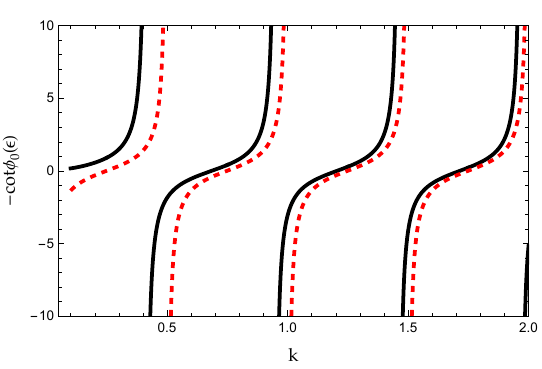}
 \caption{Coulomb effects in the generalized zeta function in Eq.\eqref{phi0}:  the Coulomb-corrected $(-\cot \phi_0)$ (solid black) vs. its non-Coulomb-limit  (dashed red)  at trap radius $R=2\pi$.  The reduced mass is taken as $\mu=1$ and Coulomb potential strength as $Z=0.1$.  The pole position at non-Coulomb-limit is simply given by $k_n = \frac{n \pi}{R} = \frac{}n{2}$, where $n\in [0, 1, \cdots]$. With Coulomb correction, pole positions are shifted by Coulomb potential, and must be numerically solved by Eq.(\ref{qcCoulomb}). \label{phi0plot} }
 \end{figure}

  \begin{figure}
\includegraphics[width=0.99\textwidth]{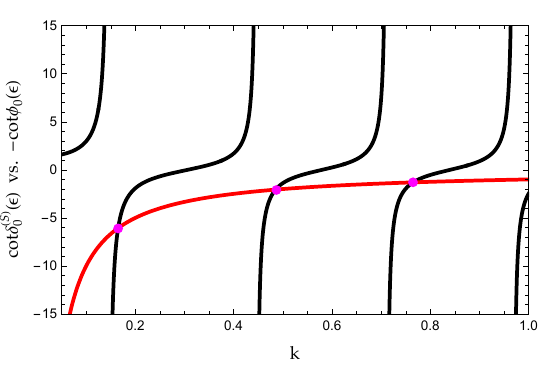}
 \caption{Quantization condition plot of Eq.\eqref{qcVR} for two particles interacting through a contact potential of strength $V_0$ and Coulomb  potential of strength $Z$ in a spherical hard-wall trap of radius $R$. The black curves correspond to the Coulomb-corrected `trapped phase angle' $(\phi_0)$ and the red curve to the infinite volume phase shift $ ( \delta_0^{(S)}) $.  The quantized energy levels are at the intersection points (purple dots) of black and red curves. The parameters are taken as:  $V_R=0.5$, $Z=0.1$, $R=4\pi$, $\mu=1$.   \label{qcplot} }
 \end{figure}

  \begin{figure}
\includegraphics[width=0.99\textwidth]{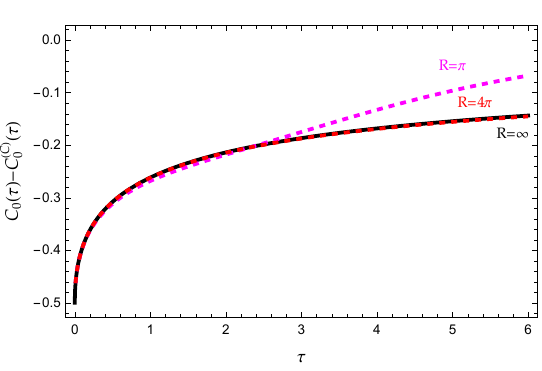}
 \caption{ Convergence of Eq.\eqref{numericeq}. The difference of integrated correlation functions (dashed red)  at two values of trap radius  $R=\pi$ (dashed purple), $4\pi$ (dashed red) vs. the result of a weighted integral from infinite volume phase shifts (solid black).  The rest of parameters are taken as:  $V_R=0.5$, $\mu=1$, and $Z=0.1$. 
 \label{ctplot} }
 \end{figure}

The effects of Coulomb interaction in the generalized zeta function are demonstrated in Fig.~\ref{phi0plot}. We see that Coulomb corrections can be fairly significant at low energies, and become smaller as energy is increased.

In Fig.~\ref{qcplot}, we show how the quantization condition in Eq.\eqref{qcVR}, or equivalently Eq.\eqref{qchsw},  works. The $\cot  \delta_0^{(S)}$  from the infinite volume phase shift in Eq.\eqref{phase} is plotted as a function of CM momentum $k$ (red curve). On the same graph the generalized zeta function  $(- \cot \phi_0  )$ from Eq.\eqref{phi0} is also plotted (black curves). Where the two intercept gives rise to  quantized energy levels in the system. Each quantized energy level is found between two neighboring poles of the generalized zeta function which correspond to  the non-interacting levels in the trap. This is to be expected because interaction only causes  the non-interacting levels to shift. The amount of energy shift is characterized by the infinite volume phase shifts.

Finally, we show in Fig.~\ref{ctplot} how the Coulomb-modified formula in Eq.\eqref{numericeq} is satisfied. 
The right-hand side from the infinite volume phase shift is plotted as a function of Euclidean time $\tau$ (black curve).
The left-hand side is obtained from summing over two energy spectra $\epsilon_n$ and $\epsilon^{(C)}_n$. The formula requires the limit of infinite trap size.  So we evaluate it at two values of the hard-wall radius $R$ (colored curves).   About a hundred energy levels are required to saturated the sum.
We see a rapid convergence of the relation over a wide range of  times as the trap size is increased. Moreover, the convergence is faster at earlier times.

\section{Summary and outlook}\label{sec:summary}

The integrated correlation function formalism developed for scattering in Refs.~\cite{Guo:2023ecc,Guo:2024zal,Guo:2024pvt}  is now extended to include  long-range Coulomb interaction for charged particle systems. The central result is the Coulomb-modified relation in Eq.\eqref{CtpartialwavSC} or its Euclidean counterpart in Eq.\eqref{CtpartialwavSCE}. Remarkably, it retains the same form as the original Eq.(\ref{Ctpartialwav}) in the absence of Coulomb interaction, but with a new meaning. On the left-hand side, instead of the difference between interacting and non-interacting integrated correlation functions for a short-range potential, it is the difference between short-range plus Coulomb potential and pure Coulomb potential. On the right-hand side, it involves the infinite volume phase shift due to short-range potential only. The realization comes from a careful examination of the asymptotic behavior of the formalism in the presence of Coulomb interaction.
The new relation is free of cutoff dependence and divergences, with a well-defined approach to the infinite volume limit.  The relation is verified to high precision with an exactly solvable model: a contact interaction potential and a spherical hard-wall trap. Several additional comments are in order.

The relation is given in 3+1 dimensions in a partial-wave projection. 
It is valid for any short-range interaction that can be described by a central potential.
The trap used to contain the system can either have continuous rotational symmetry (hard-wall, harmonic oscillator, etc), or discrete symmetry (periodic boxes/lattices) in which case a projection to the irreducible representations of the symmetry group is required.

It is worth emphasizing two properties of the integrated correlation function formalism, in comparison to L\"{u}scher-like quantization conditions:
a) it only requires correlation functions, not energy spectrum (even though an energy spectral representation can be used as in our numerical verification);
b) it has faster convergence at smaller Euclidean times.
Both properties bode well for lattice QCD simulations of multi-baryon systems where signal-to-noise ratio and energy spectrum 
extraction present major challenges, especially when large  volumes are involved.

In practical applications, the extraction of phase shift from the integrated correlation function formalism is essentially an {\em inverse problem} for which there are well-established methods in the literature, see e.g. Refs.~\cite{Rothkopf_2017,Rothkopf_2022,yang2023}.

Although the derivation is based on non-relativistic dynamics, we envision no conceptual issues extending it to relativistic dynamics \cite{Guo:2024zal}.

\acknowledgments
This research is supported by the U.S. National Science Foundation under grant PHY-2418937 (P.G.) and in part PHY-1748958 (P.G.), and the U.S. Department of Energy under grant  DE-FG02-95ER40907 (F.L.).

\appendix

\section{Scattering solutions with a contact interaction in infinite volume}\label{scatteringsolutions}  
In infinite volume,  with a incoming plane wave of $e^{i \mathbf{ k} \cdot \mathbf{ r}} $,    
 the scattering solution of two particles interaction is described by inhomogeneous  integral   Lippmann-Schwinger (LS) equation,  
\begin{align}
 & \psi^{(\infty)}_{ \epsilon}(\mathbf{ r},\mathbf{ k})  =e^{i \mathbf{ k} \cdot \mathbf{ r}}   \nonumber \\
 & + \int_{ -\infty}^\infty d \mathbf{ r}' G^{(0, \infty)} (\mathbf{ r}- \mathbf{ r}' ; \epsilon )   V_S ( \mathbf{ r'})   \psi^{(\infty)}_{ \epsilon}(\mathbf{ r}',\mathbf{ k})    , 
\end{align}  
where $k = \sqrt{2\mu \epsilon}$, and the non-interacting Green's function is given by
\begin{equation}
 G^{(0, \infty)} (\mathbf{ r}   ; \epsilon)    = \int \frac{d \mathbf{ p}}{(2\pi)^3} \frac{ e^{i \mathbf{ p} \cdot  \mathbf{ r}  }  }{\frac{q^2}{2\mu} - \frac{\mathbf{ p}^2}{2\mu}}   = - \frac{2\mu }{4\pi} i  k h_0^{(+)} (k | \mathbf{ r}   |).
\end{equation}

With a contact interaction $V_S(\mathbf{ r}) = V_0 \frac{ \delta(r)}{r^2}$,  LS equation is reduced to
\begin{equation}
  \psi^{(\infty)}_{ \epsilon}(\mathbf{ r},\mathbf{ k})  =e^{i \mathbf{ k} \cdot \mathbf{ r}}      - i   2\mu k V_0 h_0^{(+)} (k  r )     \psi^{(\infty)}_{ \epsilon}( 0,\mathbf{ k})    .
\end{equation}  
Hence only $S$-wave contribute to scattering solutions: 
\begin{equation}
  \psi^{(\infty)}_{ 0}(r; \epsilon)  =4\pi j_0 (kr)     - i   2\mu k V_0 h_0^{(+)} (k  r )     \psi^{(\infty)}_{ 0}(0; \epsilon)   ,
\end{equation}  
where the partial wave expansion is defined by
\begin{equation}
 \psi^{(\infty)}_{ \epsilon }(\mathbf{ r},\mathbf{ k})  = \sum_{l m_l}  Y^*_{lm_l} (\mathbf{ \hat{k}}) \psi^{(\infty)}_{  l}(r; \epsilon) Y_{lm_l} (\mathbf{ \hat{r}}).
\end{equation}
 The formal solution is thus given by
\begin{equation}
  \psi^{(\infty)}_{ 0}(r; \epsilon) =4\pi j_0 (kr)      +  4\pi   i  t_0^{(S)} (\epsilon )  h_0^{(+)} (k  r )       , 
\end{equation} 
where the scattering amplitude is defined by
\begin{equation}
t^{(S)}_0(\epsilon) =  - \frac{   2\mu k    }{\frac{1}{V_0} +  i   2\mu k     h_0^{(+)} (k  r ) |_{r\rightarrow 0}  } .
\end{equation}
Using asymptotic behavior of spherical Hankel function: $h_0^{(+)} (k  r ) |_{r\rightarrow 0}  \rightarrow  1 - \frac{i}{k r}$, and also introducing the renormalized coupling strength by  
\begin{equation}
    V_R = \left ( \frac{1}{V_0} +   \frac{2\mu   }{  r}|_{r\rightarrow 0} \right )^{-1},
    \label{VR}
    \end{equation} 
    the scattering amplitude now is given by
\begin{equation}
t^{(S)}_0(\epsilon) = -  \frac{  1   }{\frac{1 }{ 2\mu  V_R k} +  i        } .
\end{equation}
The scattering phase shift due to contact interaction is hence given by
\begin{equation}
\cot \delta_0^{(S)} (\epsilon)  = i + \frac{1}{t^{(S)}_0(\epsilon) } =  - \frac{1 }{ 2\mu  V_R k}.
\end{equation}

\section{Connecting eigensolutions in a trap to infinite volume scattering solutions with  Coulomb interaction}\label{traptoscattappend}
In this section, we discuss some exact solutions involving Coulomb interaction.
They are needed in the validation of the new relation in Eq.\eqref{CtpartialwavSC}.

\subsection{Exact solution of partial-wave-projected  Green's function for pure Coulomb interaction}
In addition to the asymptotic wavefunction in Eq.\eqref{SCwavasymp} and the phase shift in Eq.\eqref{Coulombphaseshift},
the Green's function also has an analytical expression for Coulomb interaction. The Green's function for $l$-th partial wave is the solution of differential equation,
\begin{align}
& \left  [ \epsilon +  \frac{1}{2\mu r^2} \frac{d}{d r} \left ( r^2 \frac{d }{d r}  \right ) - \frac{l (l+1)}{ 2\mu r^2}  - V_C (r)    \right ]G_l^{(C,\infty)} (  r,r'; \epsilon)  \nonumber \\
& = \frac{ \delta(r-r')  }{r^2},
\end{align}
whose analytical expression is given by~\cite{messiah1999quantum,Hostler1964,Guo:2021qfu},
 \begin{equation}
G_{ l }^{(C,\infty)} (r, r' ; \epsilon)    = - i 2 \mu  k j_l^{(C)} ( \gamma, k r_<)   h_l^{(C,+)} (\gamma, k r_>)  , \label{Gcoulombinf}
\end{equation}
where $r_<$ and $r_>$ represent the lesser and greater of $(r,r')$ respectively,  and
\begin{equation}
  h_l^{(C,\pm)} (\gamma, k r) =  j_l^{(C)} (\gamma, k r)  \pm i  n_l^{(C)} (\gamma, k r)   .
\end{equation}
 The Coulomb-modified  spherical Bessel functions $j_l^{(C)}(\gamma, k r)$ and $n_l^{(C)}(\gamma, k r) $ are defined~\cite{Guo:2021qfu}  via two linearly independent Kummer functions  $M(a,b,z) $ and $U(a,b,z)$, 
\begin{align}
& j_l^{(C)}(\gamma, k r)= \mathbb{C}_l (\gamma)     (k r)^l  e^{i k r} M(l+1+ i \gamma, 2l+2, -2 i k r), \label{jLC}
\end{align}
and
\begin{align}
& n_l^{(C)}(\gamma, k r)  \nonumber \\
&=  i(-2k r)^l e^{ \frac{\pi}{2}\gamma}     e^{i k r} U(l+1+ i \gamma, 2l+2, -2 i k r) e^{ i \delta_l^{(C)}}    \nonumber \\
 &-i(-2k r)^l e^{ \frac{\pi}{2}\gamma}      e^{-i k r} U(l+1- i \gamma, 2l+2, 2 i k r) e^{- i \delta_l^{(C)}}  , \label{nLC}
\end{align}
The Sommerfeld factor  is defined by \cite{messiah1999quantum},
\begin{equation}
\mathbb{C}_l (\gamma) = 2^l \frac{|\Gamma(l+1+ i \gamma)|}{(2 l+1)!} e^{- \frac{\pi}{2} \gamma }.
\end{equation}
 The Coulomb phase shift, $\delta_l^{(C)}$, is defined in Eq.(\ref{Coulombphaseshift}).

\subsection{Coulomb-modified scattering solutions with a contact interaction in infinite volume}\label{Coulombscatteringsolutions}  
In infinite volume (no trap),    the scattering solution of two charged particles interacting via a  Coulomb potential is described by the inhomogeneous LS equation,  see e.g. Ref.~\cite{Guo:2021qfu},
\begin{align}
 & \psi^{(\infty)}_{ \epsilon}(\mathbf{ r},\mathbf{ k})  =\psi^{(C,\infty)}_{ \epsilon}(\mathbf{ r},\mathbf{ k})    \nonumber \\
 & + \int_{ -\infty}^\infty d \mathbf{ r}' G^{(C, \infty)} (\mathbf{ r}, \mathbf{ r}' ; \epsilon )   V_S ( \mathbf{ r'})   \psi^{(\infty)}_{ \epsilon}(\mathbf{ r}',\mathbf{ k})    , 
\end{align}  
where $\psi^{(C,\infty)}_{ \epsilon}(\mathbf{ r},\mathbf{ k}) $ is Coulomb wavefunction  and $  G^{(C, \infty)} (\mathbf{ r} , \mathbf{ r'}  ; \epsilon)  $ is Coulomb Green's function. 

In the case of a contact interaction $V_S(\mathbf{ r}) = V_0 \frac{\delta(r)}{r^2}$, only $S$-wave will contributes, so after partial-wave projection, we find
 \begin{equation}
  \psi^{(\infty)}_{ 0}(r; \epsilon)  =\psi^{(C,\infty)}_{ 0}(r; \epsilon)   +  V_0    G_0^{(C, \infty)} (r, 0 ; \epsilon ) \psi^{(\infty)}_{ 0}(0; \epsilon)   , 
\end{equation}  
where the analytic expression of $S$-wave Coulomb  wave function and Coulomb Green's function are given by
\begin{equation}
\psi^{(C,\infty)}_{ 0}(r; \epsilon)  = 4\pi j_0^{(C)} (\gamma, k r) e^{  i \delta_0^{(C)} (\epsilon) } , 
\end{equation}
and 
\begin{equation}
G_0^{(C, \infty)} (r, r' ; \epsilon ) = - i 2\mu k j_0^{(C)} (\gamma, k r_<)   h_0^{(C, +)} (\gamma, k r_>) .
\end{equation}

The formal scattering solution in presence of Coulomb force can thus be obtained,
 \begin{align}
  \psi^{(\infty)}_{ 0}(r; \epsilon)  &=  4\pi  j_0^{(C)} (\gamma, k r) e^{  i \delta_0^{(C)} (\epsilon) }  \nonumber \\
&  +  4\pi   i t_0^{(SC)}  (\epsilon ) h_0^{(C, +)} (\gamma, k r)  e^{ -  i \delta_0^{(C)} (\epsilon) }   , 
\end{align} 
 where 
\begin{equation}
t_0^{(SC)} (\epsilon) =    \frac{  - 2\mu k    \mathbb{C}^2_0 (\gamma) e^{  2 i \delta_0^{(C)} (\epsilon) }  }{\frac{1}{ V_0} -    Re \left [  G_0^{(C, \infty)} (r, 0 ; \epsilon )|_{r\rightarrow 0} \right ] +  i 2 \mu k  \mathbb{C}^2_0 (\gamma)}  .
\end{equation}
The $t_0^{(SC)} (\epsilon) $ is usually parameterized by
\begin{equation}
t_0^{(SC)} (\epsilon)  = \frac{1 }{\cot \delta_0^{(S) } (\epsilon) - i } e^{  2 i \delta_0^{(C)} (\epsilon) } .
\end{equation}
Hence we find a useful relation
\begin{equation}
  \frac{1}{ V_0} =  - 2\mu k    \mathbb{C}^2_0 (\gamma)  \cot \delta_0^{(S)}  (\epsilon) +Re \left [  G_0^{(C, \infty)} (r, 0 ; \epsilon )|_{r\rightarrow 0} \right ]       . \label{V0scattrelation}
\end{equation}
The total scattering amplitude is defined by
\begin{equation}
t_0(\epsilon) = t^{(C)}_0(\epsilon)  + t^{(SC)}_0(\epsilon)  = \frac{e^{2 i \delta_0 (\epsilon)} -1}{2i},
\end{equation}
where $ t^{(C)}_0(\epsilon) $ is pure Coulomb scattering amplitude
\begin{equation}
t_0^{(C)} (\epsilon)  = \frac{e^{2 i \delta^{(C)}_0 (\epsilon)} -1}{2i}  = \frac{1 }{\cot \delta_0^{(C) } (\epsilon) - i }, 
\end{equation}
and $\delta_0 (\epsilon)$ is total $S$-wave scattering amplitude,
\begin{equation}
\delta_0 (\epsilon) = \delta^{(S)}_0 (\epsilon) + \delta^{(C)}_0 (\epsilon).
\end{equation}

With Coulomb correction, the Friedel formula and Krein's theorem in Eq.(\ref{Friedelformula})   must be replaced by,
 \begin{equation} 
  Tr  \left [ G_l^{(\infty)} (  \epsilon) -G_l^{(C, \infty)} (    \epsilon )   \right ]  = - \frac{1}{\pi}  \int_{0}^{\infty} d \lambda  \frac{\delta^{(S)}_l  (\lambda)}{ (\lambda - \epsilon)^2}, \label{FriedelformulaCoulomb}
 \end{equation}
 where $G_l^{(\infty)} (  \epsilon) $ and $G_l^{(C, \infty)} (    \epsilon ) $ represent Green's function operators with short-range plus Coulomb interaction and  only Coulomb interaction respectively. The  infinite volume Green's function, $G_l^{(\infty)} (  r,r'; \epsilon) $, satisfies Dyson   equation,
 \begin{align}
 &  G_l^{(\infty)} (  r,r'; \epsilon)  =G_l^{(C,\infty)} (  r,r'; \epsilon)   \nonumber \\
 & + \int_0^\infty {r''}^2 d r'' G_l^{(C,\infty)} (  r,r''; \epsilon)  V_S(r'') G_l^{(\infty)} (  r'',r'; \epsilon).
 \end{align}

\subsection{Coulomb-modified quantization condition in spherical hard-wall trap with a contact interaction}\label{qctrapappend}
The    dynamics of particles system in a   trap    can be also described    by the homogeneous Lippmann-Schwinger equation,
\begin{equation}
\psi (\mathbf{ r}; \epsilon) = \int  d \mathbf{ r'} G^{(C, trap)} (\mathbf{ r},\mathbf{ r'}; \epsilon) V_S(\mathbf{ r'}) \psi (\mathbf{ r'} ; \epsilon) , \label{LStrap}
\end{equation}
where  $G^{(C, trap)}  (\mathbf{ r},\mathbf{ r'}; \epsilon) $ is Coulomb-modified Green's function for the trapped system, and it satisfies differential equation,
\begin{equation}
 \left  [ \epsilon +  \frac{\nabla^2}{2\mu }    -U_{trap} (r) - V_C (r)    \right ] G^{(C,trap)} (  \mathbf{ r},\mathbf{ r'}; \epsilon)    =   \delta(\mathbf{ r}-\mathbf{ r'})  .
\end{equation}
The partial wave expansion of Green's function is defined by
\begin{equation}
G^{(C,trap)} (  \mathbf{ r},\mathbf{ r'}; \epsilon)   = \sum_{ l m_l} Y_{lm_l}(\mathbf{ \hat{r}})  G_l^{(C,trap)} (  r,r'; \epsilon)Y_{lm_l}^*(\mathbf{ \hat{r'}})  .
\end{equation}
For a spherical hard-wall trap, the analytical form of  Coulomb-modified partial wave Green's function for the trapped system can be obtained, see Eq.(73) in Ref.\cite{Guo:2021qfu},
\begin{align}
G_l^{(C,trap)} (  r,r'; \epsilon) & = - 2 \mu k j_l^{(C)} (\gamma , k r_<)  j_l^{(C)} (\gamma , k r_>) \nonumber \\
&  \times \left [ \frac{ n_l^{(C)} (\gamma , k R) }{ j_l^{(C)} (\gamma , k R) } - \frac{ n_l^{(C)} (\gamma , k r_>) }{ j_l^{(C)} (\gamma , k r_>) } \right ], \label{GCLtrap}
\end{align}
where    the Coulomb-modified  $j_l^{(C)}(\gamma, k r)$ and $n_l^{(C)}(\gamma, k r) $ functions  are defined in Eq.(\ref{jLC}) and Eq.(\ref{nLC}) respectively.

With a contact interaction, $V_S(\mathbf{ r}) = V_0 \frac{\delta(r)}{r^2}$,  after the $S$-wave projection, Eq.(\ref{LStrap}) yields a quantization condition,
\begin{equation}
\frac{1}{V_0} = G_0^{(C,trap)} (  r, 0; \epsilon) |_{r\rightarrow 0} .
\end{equation} 
Using relation in Eq.(\ref{V0scattrelation}), we find
\begin{align}
&    \cot \delta_0^{(S)} (\epsilon)  \nonumber\\  
&= \frac{ Re \left [  G_0^{(C, \infty)} (r, 0 ; \epsilon )|_{r\rightarrow 0} \right ] -G_0^{(C,trap)} (  r, 0; \epsilon) |_{r\rightarrow 0}  }{2\mu k  \mathbb{C}^2_0 (\gamma)  } .
\end{align} 
The real part of Coulomb Green's function in infinite volume is given by
\begin{equation}
Re\left [G_0^{(C, \infty)} (r, r' ; \epsilon ) \right ] =   2\mu k j_0^{(C)} (\gamma, k r_<)   n_0^{(C)} (\gamma, k r_>) ,
\end{equation}
which cancel out the second term of   Coulomb Green's function in the trap in Eq.(\ref{GCLtrap}).
Also using asymptotic form
\begin{equation}
j_0^{(C)} (\gamma , k r) \stackrel{r\rightarrow 0}{\rightarrow} \mathbb{C}_0 (\gamma),
\end{equation}
the Coulomb-modified quantization condition for a contact interaction potential is found in Eq.\eqref{qchsw}.

\bibliography{ALL-REF.bib}

\begin{thebibliography}{73}%
\makeatletter
\providecommand \@ifxundefined [1]{%
 \@ifx{#1\undefined}
}%
\providecommand \@ifnum [1]{%
 \ifnum #1\expandafter \@firstoftwo
 \else \expandafter \@secondoftwo
 \fi
}%
\providecommand \@ifx [1]{%
 \ifx #1\expandafter \@firstoftwo
 \else \expandafter \@secondoftwo
 \fi
}%
\providecommand \natexlab [1]{#1}%
\providecommand \enquote  [1]{``#1''}%
\providecommand \bibnamefont  [1]{#1}%
\providecommand \bibfnamefont [1]{#1}%
\providecommand \citenamefont [1]{#1}%
\providecommand \href@noop [0]{\@secondoftwo}%
\providecommand \href [0]{\begingroup \@sanitize@url \@href}%
\providecommand \@href[1]{\@@startlink{#1}\@@href}%
\providecommand \@@href[1]{\endgroup#1\@@endlink}%
\providecommand \@sanitize@url [0]{\catcode `\\12\catcode `\$12\catcode
  `\&12\catcode `\#12\catcode `\^12\catcode `\_12\catcode `\%12\relax}%
\providecommand \@@startlink[1]{}%
\providecommand \@@endlink[0]{}%
\providecommand \url  [0]{\begingroup\@sanitize@url \@url }%
\providecommand \@url [1]{\endgroup\@href {#1}{\urlprefix }}%
\providecommand \urlprefix  [0]{URL }%
\providecommand \Eprint [0]{\href }%
\providecommand \doibase [0]{http://dx.doi.org/}%
\providecommand \selectlanguage [0]{\@gobble}%
\providecommand \bibinfo  [0]{\@secondoftwo}%
\providecommand \bibfield  [0]{\@secondoftwo}%
\providecommand \translation [1]{[#1]}%
\providecommand \BibitemOpen [0]{}%
\providecommand \bibitemStop [0]{}%
\providecommand \bibitemNoStop [0]{.\EOS\space}%
\providecommand \EOS [0]{\spacefactor3000\relax}%
\providecommand \BibitemShut  [1]{\csname bibitem#1\endcsname}%
\let\auto@bib@innerbib\@empty
\bibitem [{\citenamefont {Guo}\ and\ \citenamefont
  {Gasparian}(2023)}]{Guo:2023ecc}%
  \BibitemOpen
  \bibfield  {author} {\bibinfo {author} {\bibfnamefont {P.}~\bibnamefont
  {Guo}}\ and\ \bibinfo {author} {\bibfnamefont {V.}~\bibnamefont
  {Gasparian}},\ }\bibfield  {title} {\enquote {\bibinfo {title} {{Toward
  extracting the scattering phase shift from integrated correlation
  functions}},}\ }\href {\doibase 10.1103/PhysRevD.108.074504} {\bibfield
  {journal} {\bibinfo  {journal} {Phys. Rev. D}\ }\textbf {\bibinfo {volume}
  {108}},\ \bibinfo {pages} {074504} (\bibinfo {year} {2023})},\ \Eprint
  {http://arxiv.org/abs/2307.12951}{arXiv:2307.12951 [hep-lat]}\BibitemShut
  {NoStop}%
\bibitem [{\citenamefont {Guo}(2024)}]{Guo:2024zal}%
  \BibitemOpen
  \bibfield  {author} {\bibinfo {author} {\bibfnamefont {P.}~\bibnamefont
  {Guo}},\ }\bibfield  {title} {\enquote {\bibinfo {title} {{Toward extracting
  the scattering phase shift from integrated correlation functions. II. A
  relativistic lattice field theory model}},}\ }\href {\doibase
  10.1103/PhysRevD.110.014504} {\bibfield  {journal} {\bibinfo  {journal}
  {Phys. Rev. D}\ }\textbf {\bibinfo {volume} {110}},\ \bibinfo {pages}
  {014504} (\bibinfo {year} {2024})},\ \Eprint
  {http://arxiv.org/abs/2402.15628}{arXiv:2402.15628 [hep-lat]}\BibitemShut
  {NoStop}%
\bibitem [{\citenamefont {Guo}\ and\ \citenamefont {Lee}(2025)}]{Guo:2024pvt}%
  \BibitemOpen
  \bibfield  {author} {\bibinfo {author} {\bibfnamefont {P.}~\bibnamefont
  {Guo}}\ and\ \bibinfo {author} {\bibfnamefont {F.~X.}\ \bibnamefont {Lee}},\
  }\bibfield  {title} {\enquote {\bibinfo {title} {{Toward extracting
  scattering phase shift from integrated correlation functions. III. Coupled
  channels}},}\ }\href {\doibase 10.1103/PhysRevD.111.054506} {\bibfield
  {journal} {\bibinfo  {journal} {Phys. Rev. D}\ }\textbf {\bibinfo {volume}
  {111}},\ \bibinfo {pages} {054506} (\bibinfo {year} {2025})},\ \Eprint
  {http://arxiv.org/abs/2412.00812}{arXiv:2412.00812 [hep-lat]}\BibitemShut
  {NoStop}%
\bibitem [{\citenamefont {L{\"u}scher}(1991)}]{Luscher:1990ux}%
  \BibitemOpen
  \bibfield  {author} {\bibinfo {author} {\bibfnamefont {M.}~\bibnamefont
  {L{\"u}scher}},\ }\bibfield  {title} {\enquote {\bibinfo {title} {{Two
  particle states on a torus and their relation to the scattering matrix}},}\
  }\href {\doibase 10.1016/0550-3213(91)90366-6} {\bibfield  {journal}
  {\bibinfo  {journal} {Nucl. Phys.}\ }\textbf {\bibinfo {volume} {B354}},\
  \bibinfo {pages} {531} (\bibinfo {year} {1991})}\BibitemShut {NoStop}%
\bibitem [{\citenamefont {Busch}\ \emph {et~al.}(1998)\citenamefont {Busch},
  \citenamefont {Englert}, \citenamefont {Rza\.zewski},\ and\ \citenamefont
  {Wilkens}}]{Busch98}%
  \BibitemOpen
  \bibfield  {author} {\bibinfo {author} {\bibfnamefont {T.}~\bibnamefont
  {Busch}}, \bibinfo {author} {\bibfnamefont {B.-G.}\ \bibnamefont {Englert}},
  \bibinfo {author} {\bibfnamefont {K.}~\bibnamefont {Rza\.zewski}}, \ and\
  \bibinfo {author} {\bibfnamefont {M.}~\bibnamefont {Wilkens}},\ }\bibfield
  {title} {\enquote {\bibinfo {title} {{Two Cold Atoms in a Harmonic Trap}},}\
  }\href {\doibase 10.1023/A:1018705520999} {\bibfield  {journal} {\bibinfo
  {journal} {Found. Phys.}\ }\textbf {\bibinfo {volume} {28}},\ \bibinfo
  {pages} {549–559} (\bibinfo {year} {1998})}\BibitemShut {NoStop}%
\bibitem [{\citenamefont {Rummukainen}\ and\ \citenamefont
  {Gottlieb}(1995)}]{Rummukainen:1995vs}%
  \BibitemOpen
  \bibfield  {author} {\bibinfo {author} {\bibfnamefont {K.}~\bibnamefont
  {Rummukainen}}\ and\ \bibinfo {author} {\bibfnamefont {S.~A.}\ \bibnamefont
  {Gottlieb}},\ }\bibfield  {title} {\enquote {\bibinfo {title} {{Resonance
  scattering phase shifts on a nonrest frame lattice}},}\ }\href {\doibase
  10.1016/0550-3213(95)00313-H} {\bibfield  {journal} {\bibinfo  {journal}
  {Nucl. Phys.}\ }\textbf {\bibinfo {volume} {B450}},\ \bibinfo {pages} {397}
  (\bibinfo {year} {1995})},\ \Eprint
  {http://arxiv.org/abs/hep-lat/9503028}{arXiv:hep-lat/9503028
  [hep-lat]}\BibitemShut {NoStop}%
\bibitem [{\citenamefont {Christ}\ \emph {et~al.}(2005)\citenamefont {Christ},
  \citenamefont {Kim},\ and\ \citenamefont {Yamazaki}}]{Christ:2005gi}%
  \BibitemOpen
  \bibfield  {author} {\bibinfo {author} {\bibfnamefont {N.~H.}\ \bibnamefont
  {Christ}}, \bibinfo {author} {\bibfnamefont {C.}~\bibnamefont {Kim}}, \ and\
  \bibinfo {author} {\bibfnamefont {T.}~\bibnamefont {Yamazaki}},\ }\bibfield
  {title} {\enquote {\bibinfo {title} {{Finite volume corrections to the
  two-particle decay of states with non-zero momentum}},}\ }\href {\doibase
  10.1103/PhysRevD.72.114506} {\bibfield  {journal} {\bibinfo  {journal} {Phys.
  Rev.}\ }\textbf {\bibinfo {volume} {D72}},\ \bibinfo {pages} {114506}
  (\bibinfo {year} {2005})},\ \Eprint
  {http://arxiv.org/abs/hep-lat/0507009}{arXiv:hep-lat/0507009
  [hep-lat]}\BibitemShut {NoStop}%
\bibitem [{\citenamefont {Bernard}\ \emph {et~al.}(2008)\citenamefont
  {Bernard}, \citenamefont {Lage}, \citenamefont {Mei{\ss}ner},\ and\
  \citenamefont {Rusetsky}}]{Bernard:2008ax}%
  \BibitemOpen
  \bibfield  {author} {\bibinfo {author} {\bibfnamefont {V.}~\bibnamefont
  {Bernard}}, \bibinfo {author} {\bibfnamefont {M.}~\bibnamefont {Lage}},
  \bibinfo {author} {\bibfnamefont {U.-G.}\ \bibnamefont {Mei{\ss}ner}}, \ and\
  \bibinfo {author} {\bibfnamefont {A.}~\bibnamefont {Rusetsky}},\ }\bibfield
  {title} {\enquote {\bibinfo {title} {{Resonance properties from the
  finite-volume energy spectrum}},}\ }\href {\doibase
  10.1088/1126-6708/2008/08/024} {\bibfield  {journal} {\bibinfo  {journal}
  {JHEP}\ }\textbf {\bibinfo {volume} {08}},\ \bibinfo {pages} {024} (\bibinfo
  {year} {2008})},\ \Eprint {http://arxiv.org/abs/0806.4495}{arXiv:0806.4495
  [hep-lat]}\BibitemShut {NoStop}%
\bibitem [{\citenamefont {He}\ \emph {et~al.}(2005)\citenamefont {He},
  \citenamefont {Feng},\ and\ \citenamefont {Liu}}]{He:2005ey}%
  \BibitemOpen
  \bibfield  {author} {\bibinfo {author} {\bibfnamefont {S.}~\bibnamefont
  {He}}, \bibinfo {author} {\bibfnamefont {X.}~\bibnamefont {Feng}}, \ and\
  \bibinfo {author} {\bibfnamefont {C.}~\bibnamefont {Liu}},\ }\bibfield
  {title} {\enquote {\bibinfo {title} {{Two particle states and the S-matrix
  elements in multi-channel scattering}},}\ }\href {\doibase
  10.1088/1126-6708/2005/07/011} {\bibfield  {journal} {\bibinfo  {journal}
  {JHEP}\ }\textbf {\bibinfo {volume} {07}},\ \bibinfo {pages} {011} (\bibinfo
  {year} {2005})},\ \Eprint
  {http://arxiv.org/abs/hep-lat/0504019}{arXiv:hep-lat/0504019
  [hep-lat]}\BibitemShut {NoStop}%
\bibitem [{\citenamefont {Lage}\ \emph {et~al.}(2009)\citenamefont {Lage},
  \citenamefont {Mei{\ss}ner},\ and\ \citenamefont {Rusetsky}}]{Lage:2009zv}%
  \BibitemOpen
  \bibfield  {author} {\bibinfo {author} {\bibfnamefont {M.}~\bibnamefont
  {Lage}}, \bibinfo {author} {\bibfnamefont {U.-G.}\ \bibnamefont
  {Mei{\ss}ner}}, \ and\ \bibinfo {author} {\bibfnamefont {A.}~\bibnamefont
  {Rusetsky}},\ }\bibfield  {title} {\enquote {\bibinfo {title} {{A Method to
  measure the antikaon-nucleon scattering length in lattice QCD}},}\ }\href
  {\doibase 10.1016/j.physletb.2009.10.055} {\bibfield  {journal} {\bibinfo
  {journal} {Phys. Lett.}\ }\textbf {\bibinfo {volume} {B681}},\ \bibinfo
  {pages} {439} (\bibinfo {year} {2009})},\ \Eprint
  {http://arxiv.org/abs/0905.0069}{arXiv:0905.0069 [hep-lat]}\BibitemShut
  {NoStop}%
\bibitem [{\citenamefont {D{\"o}ring}\ \emph {et~al.}(2011)\citenamefont
  {D{\"o}ring}, \citenamefont {Mei{\ss}ner}, \citenamefont {Oset},\ and\
  \citenamefont {Rusetsky}}]{Doring:2011vk}%
  \BibitemOpen
  \bibfield  {author} {\bibinfo {author} {\bibfnamefont {M.}~\bibnamefont
  {D{\"o}ring}}, \bibinfo {author} {\bibfnamefont {U.-G.}\ \bibnamefont
  {Mei{\ss}ner}}, \bibinfo {author} {\bibfnamefont {E.}~\bibnamefont {Oset}}, \
  and\ \bibinfo {author} {\bibfnamefont {A.}~\bibnamefont {Rusetsky}},\
  }\bibfield  {title} {\enquote {\bibinfo {title} {{Unitarized Chiral
  Perturbation Theory in a finite volume: Scalar meson sector}},}\ }\href
  {\doibase 10.1140/epja/i2011-11139-7} {\bibfield  {journal} {\bibinfo
  {journal} {Eur. Phys. J.}\ }\textbf {\bibinfo {volume} {A47}},\ \bibinfo
  {pages} {139} (\bibinfo {year} {2011})},\ \Eprint
  {http://arxiv.org/abs/1107.3988}{arXiv:1107.3988 [hep-lat]}\BibitemShut
  {NoStop}%
\bibitem [{\citenamefont {Guo}\ \emph {et~al.}(2013)\citenamefont {Guo},
  \citenamefont {Dudek}, \citenamefont {Edwards},\ and\ \citenamefont
  {Szczepaniak}}]{Guo:2012hv}%
  \BibitemOpen
  \bibfield  {author} {\bibinfo {author} {\bibfnamefont {P.}~\bibnamefont
  {Guo}}, \bibinfo {author} {\bibfnamefont {J.}~\bibnamefont {Dudek}}, \bibinfo
  {author} {\bibfnamefont {R.}~\bibnamefont {Edwards}}, \ and\ \bibinfo
  {author} {\bibfnamefont {A.~P.}\ \bibnamefont {Szczepaniak}},\ }\bibfield
  {title} {\enquote {\bibinfo {title} {{Coupled-channel scattering on a
  torus}},}\ }\href {\doibase 10.1103/PhysRevD.88.014501} {\bibfield  {journal}
  {\bibinfo  {journal} {Phys. Rev.}\ }\textbf {\bibinfo {volume} {D88}},\
  \bibinfo {pages} {014501} (\bibinfo {year} {2013})},\ \Eprint
  {http://arxiv.org/abs/1211.0929}{arXiv:1211.0929 [hep-lat]}\BibitemShut
  {NoStop}%
\bibitem [{\citenamefont {Guo}(2013)}]{Guo:2013vsa}%
  \BibitemOpen
  \bibfield  {author} {\bibinfo {author} {\bibfnamefont {P.}~\bibnamefont
  {Guo}},\ }\bibfield  {title} {\enquote {\bibinfo {title} {{Coupled-channel
  scattering in 1+1 dimensional lattice model}},}\ }\href {\doibase
  10.1103/PhysRevD.88.014507} {\bibfield  {journal} {\bibinfo  {journal} {Phys.
  Rev.}\ }\textbf {\bibinfo {volume} {D88}},\ \bibinfo {pages} {014507}
  (\bibinfo {year} {2013})},\ \Eprint
  {http://arxiv.org/abs/1304.7812}{arXiv:1304.7812 [hep-lat]}\BibitemShut
  {NoStop}%
\bibitem [{\citenamefont {Kreuzer}\ and\ \citenamefont
  {Hammer}(2009)}]{Kreuzer:2008bi}%
  \BibitemOpen
  \bibfield  {author} {\bibinfo {author} {\bibfnamefont {S.}~\bibnamefont
  {Kreuzer}}\ and\ \bibinfo {author} {\bibfnamefont {H.~W.}\ \bibnamefont
  {Hammer}},\ }\bibfield  {title} {\enquote {\bibinfo {title} {{Efimov physics
  in a finite volume}},}\ }\href {\doibase 10.1016/j.physletb.2009.02.035}
  {\bibfield  {journal} {\bibinfo  {journal} {Phys. Lett.}\ }\textbf {\bibinfo
  {volume} {B673}},\ \bibinfo {pages} {260} (\bibinfo {year} {2009})},\ \Eprint
  {http://arxiv.org/abs/0811.0159}{arXiv:0811.0159 [nucl-th]}\BibitemShut
  {NoStop}%
\bibitem [{\citenamefont {Polejaeva}\ and\ \citenamefont
  {Rusetsky}(2012)}]{Polejaeva:2012ut}%
  \BibitemOpen
  \bibfield  {author} {\bibinfo {author} {\bibfnamefont {K.}~\bibnamefont
  {Polejaeva}}\ and\ \bibinfo {author} {\bibfnamefont {A.}~\bibnamefont
  {Rusetsky}},\ }\bibfield  {title} {\enquote {\bibinfo {title} {{Three
  particles in a finite volume}},}\ }\href {\doibase
  10.1140/epja/i2012-12067-8} {\bibfield  {journal} {\bibinfo  {journal} {Eur.
  Phys. J.}\ }\textbf {\bibinfo {volume} {A48}},\ \bibinfo {pages} {67}
  (\bibinfo {year} {2012})},\ \Eprint
  {http://arxiv.org/abs/1203.1241}{arXiv:1203.1241 [hep-lat]}\BibitemShut
  {NoStop}%
\bibitem [{\citenamefont {Hansen}\ and\ \citenamefont
  {Sharpe}(2014)}]{Hansen:2014eka}%
  \BibitemOpen
  \bibfield  {author} {\bibinfo {author} {\bibfnamefont {M.~T.}\ \bibnamefont
  {Hansen}}\ and\ \bibinfo {author} {\bibfnamefont {S.~R.}\ \bibnamefont
  {Sharpe}},\ }\bibfield  {title} {\enquote {\bibinfo {title} {{Relativistic,
  model-independent, three-particle quantization condition}},}\ }\href
  {\doibase 10.1103/PhysRevD.90.116003} {\bibfield  {journal} {\bibinfo
  {journal} {Phys. Rev.}\ }\textbf {\bibinfo {volume} {D90}},\ \bibinfo {pages}
  {116003} (\bibinfo {year} {2014})},\ \Eprint
  {http://arxiv.org/abs/1408.5933}{arXiv:1408.5933 [hep-lat]}\BibitemShut
  {NoStop}%
\bibitem [{\citenamefont {Mai}\ and\ \citenamefont
  {D{\"o}ring}(2017)}]{Mai:2017bge}%
  \BibitemOpen
  \bibfield  {author} {\bibinfo {author} {\bibfnamefont {M.}~\bibnamefont
  {Mai}}\ and\ \bibinfo {author} {\bibfnamefont {M.}~\bibnamefont
  {D{\"o}ring}},\ }\bibfield  {title} {\enquote {\bibinfo {title} {{Three-body
  Unitarity in the Finite Volume}},}\ }\href {\doibase
  10.1140/epja/i2017-12440-1} {\bibfield  {journal} {\bibinfo  {journal} {Eur.
  Phys. J.}\ }\textbf {\bibinfo {volume} {A53}},\ \bibinfo {pages} {240}
  (\bibinfo {year} {2017})},\ \Eprint
  {http://arxiv.org/abs/1709.08222}{arXiv:1709.08222 [hep-lat]}\BibitemShut
  {NoStop}%
\bibitem [{\citenamefont {Mai}\ and\ \citenamefont
  {Döring}(2019)}]{Mai:2018djl}%
  \BibitemOpen
  \bibfield  {author} {\bibinfo {author} {\bibfnamefont {M.}~\bibnamefont
  {Mai}}\ and\ \bibinfo {author} {\bibfnamefont {M.}~\bibnamefont {Döring}},\
  }\bibfield  {title} {\enquote {\bibinfo {title} {{Finite-Volume Spectrum of
  $\pi^+\pi^+$ and $\pi^+\pi^+\pi^+$ Systems}},}\ }\href {\doibase
  10.1103/PhysRevLett.122.062503} {\bibfield  {journal} {\bibinfo  {journal}
  {Phys. Rev. Lett.}\ }\textbf {\bibinfo {volume} {122}},\ \bibinfo {pages}
  {062503} (\bibinfo {year} {2019})},\ \Eprint
  {http://arxiv.org/abs/1807.04746}{arXiv:1807.04746 [hep-lat]}\BibitemShut
  {NoStop}%
\bibitem [{\citenamefont {D{\"o}ring}\ \emph {et~al.}(2018)\citenamefont
  {D{\"o}ring}, \citenamefont {Hammer}, \citenamefont {Mai}, \citenamefont
  {Pang}, \citenamefont {Rusetsky},\ and\ \citenamefont {Wu}}]{Doring:2018xxx}%
  \BibitemOpen
  \bibfield  {author} {\bibinfo {author} {\bibfnamefont {M.}~\bibnamefont
  {D{\"o}ring}}, \bibinfo {author} {\bibfnamefont {H.~W.}\ \bibnamefont
  {Hammer}}, \bibinfo {author} {\bibfnamefont {M.}~\bibnamefont {Mai}},
  \bibinfo {author} {\bibfnamefont {J.~Y.}\ \bibnamefont {Pang}}, \bibinfo
  {author} {\bibfnamefont {A.}~\bibnamefont {Rusetsky}}, \ and\ \bibinfo
  {author} {\bibfnamefont {J.}~\bibnamefont {Wu}},\ }\bibfield  {title}
  {\enquote {\bibinfo {title} {{Three-body spectrum in a finite volume: the
  role of cubic symmetry}},}\ }\href {\doibase 10.1103/PhysRevD.97.114508}
  {\bibfield  {journal} {\bibinfo  {journal} {Phys. Rev.}\ }\textbf {\bibinfo
  {volume} {D97}},\ \bibinfo {pages} {114508} (\bibinfo {year} {2018})},\
  \Eprint {http://arxiv.org/abs/1802.03362}{arXiv:1802.03362
  [hep-lat]}\BibitemShut {NoStop}%
\bibitem [{\citenamefont {Guo}(2017)}]{Guo:2016fgl}%
  \BibitemOpen
  \bibfield  {author} {\bibinfo {author} {\bibfnamefont {P.}~\bibnamefont
  {Guo}},\ }\bibfield  {title} {\enquote {\bibinfo {title} {{One spatial
  dimensional finite volume three-body interaction for a short-range
  potential}},}\ }\href {\doibase 10.1103/PhysRevD.95.054508} {\bibfield
  {journal} {\bibinfo  {journal} {Phys. Rev.}\ }\textbf {\bibinfo {volume}
  {D95}},\ \bibinfo {pages} {054508} (\bibinfo {year} {2017})},\ \Eprint
  {http://arxiv.org/abs/1607.03184}{arXiv:1607.03184 [hep-lat]}\BibitemShut
  {NoStop}%
\bibitem [{\citenamefont {Guo}\ and\ \citenamefont
  {Gasparian}(2017)}]{Guo:2017ism}%
  \BibitemOpen
  \bibfield  {author} {\bibinfo {author} {\bibfnamefont {P.}~\bibnamefont
  {Guo}}\ and\ \bibinfo {author} {\bibfnamefont {V.}~\bibnamefont
  {Gasparian}},\ }\bibfield  {title} {\enquote {\bibinfo {title} {{An solvable
  three-body model in finite volume}},}\ }\href {\doibase
  10.1016/j.physletb.2017.10.009} {\bibfield  {journal} {\bibinfo  {journal}
  {Phys. Lett.}\ }\textbf {\bibinfo {volume} {B774}},\ \bibinfo {pages} {441}
  (\bibinfo {year} {2017})},\ \Eprint
  {http://arxiv.org/abs/1701.00438}{arXiv:1701.00438 [hep-lat]}\BibitemShut
  {NoStop}%
\bibitem [{\citenamefont {Guo}\ and\ \citenamefont
  {Gasparian}(2018)}]{Guo:2017crd}%
  \BibitemOpen
  \bibfield  {author} {\bibinfo {author} {\bibfnamefont {P.}~\bibnamefont
  {Guo}}\ and\ \bibinfo {author} {\bibfnamefont {V.}~\bibnamefont
  {Gasparian}},\ }\bibfield  {title} {\enquote {\bibinfo {title} {{Numerical
  approach for finite volume three-body interaction}},}\ }\href {\doibase
  10.1103/PhysRevD.97.014504} {\bibfield  {journal} {\bibinfo  {journal} {Phys.
  Rev.}\ }\textbf {\bibinfo {volume} {D97}},\ \bibinfo {pages} {014504}
  (\bibinfo {year} {2018})},\ \Eprint
  {http://arxiv.org/abs/1709.08255}{arXiv:1709.08255 [hep-lat]}\BibitemShut
  {NoStop}%
\bibitem [{\citenamefont {Guo}\ and\ \citenamefont
  {Morris}(2019)}]{Guo:2018xbv}%
  \BibitemOpen
  \bibfield  {author} {\bibinfo {author} {\bibfnamefont {P.}~\bibnamefont
  {Guo}}\ and\ \bibinfo {author} {\bibfnamefont {T.}~\bibnamefont {Morris}},\
  }\bibfield  {title} {\enquote {\bibinfo {title} {{Multiple-particle
  interaction in (1+1)-dimensional lattice model}},}\ }\href {\doibase
  10.1103/PhysRevD.99.014501} {\bibfield  {journal} {\bibinfo  {journal} {Phys.
  Rev.}\ }\textbf {\bibinfo {volume} {D99}},\ \bibinfo {pages} {014501}
  (\bibinfo {year} {2019})},\ \Eprint
  {http://arxiv.org/abs/1808.07397}{arXiv:1808.07397 [hep-lat]}\BibitemShut
  {NoStop}%
\bibitem [{\citenamefont {Mai}\ \emph {et~al.}(2020)\citenamefont {Mai},
  \citenamefont {D\"oring}, \citenamefont {Culver},\ and\ \citenamefont
  {Alexandru}}]{Mai:2019fba}%
  \BibitemOpen
  \bibfield  {author} {\bibinfo {author} {\bibfnamefont {M.}~\bibnamefont
  {Mai}}, \bibinfo {author} {\bibfnamefont {M.}~\bibnamefont {D\"oring}},
  \bibinfo {author} {\bibfnamefont {C.}~\bibnamefont {Culver}}, \ and\ \bibinfo
  {author} {\bibfnamefont {A.}~\bibnamefont {Alexandru}},\ }\bibfield  {title}
  {\enquote {\bibinfo {title} {{Three-body unitarity versus finite-volume
  $\pi^+\pi^+\pi^+$ spectrum from lattice QCD}},}\ }\href {\doibase
  10.1103/PhysRevD.101.054510} {\bibfield  {journal} {\bibinfo  {journal}
  {Phys. Rev. D}\ }\textbf {\bibinfo {volume} {101}},\ \bibinfo {pages}
  {054510} (\bibinfo {year} {2020})},\ \Eprint
  {http://arxiv.org/abs/1909.05749}{arXiv:1909.05749 [hep-lat]}\BibitemShut
  {NoStop}%
\bibitem [{\citenamefont {Guo}\ \emph {et~al.}(2018)\citenamefont {Guo},
  \citenamefont {Döring},\ and\ \citenamefont {Szczepaniak}}]{Guo:2018ibd}%
  \BibitemOpen
  \bibfield  {author} {\bibinfo {author} {\bibfnamefont {P.}~\bibnamefont
  {Guo}}, \bibinfo {author} {\bibfnamefont {M.}~\bibnamefont {Döring}}, \ and\
  \bibinfo {author} {\bibfnamefont {A.~P.}\ \bibnamefont {Szczepaniak}},\
  }\bibfield  {title} {\enquote {\bibinfo {title} {{Variational approach to
  $N$-body interactions in finite volume}},}\ }\href {\doibase
  10.1103/PhysRevD.98.094502} {\bibfield  {journal} {\bibinfo  {journal} {Phys.
  Rev.}\ }\textbf {\bibinfo {volume} {D98}},\ \bibinfo {pages} {094502}
  (\bibinfo {year} {2018})},\ \Eprint
  {http://arxiv.org/abs/1810.01261}{arXiv:1810.01261 [hep-lat]}\BibitemShut
  {NoStop}%
\bibitem [{\citenamefont {Guo}(2020{\natexlab{a}})}]{Guo:2019hih}%
  \BibitemOpen
  \bibfield  {author} {\bibinfo {author} {\bibfnamefont {P.}~\bibnamefont
  {Guo}},\ }\bibfield  {title} {\enquote {\bibinfo {title} {{Propagation of
  particles on a torus}},}\ }\href {\doibase 10.1016/j.physletb.2020.135370}
  {\bibfield  {journal} {\bibinfo  {journal} {Phys. Lett. B}\ }\textbf
  {\bibinfo {volume} {804}},\ \bibinfo {pages} {135370} (\bibinfo {year}
  {2020}{\natexlab{a}})},\ \Eprint
  {http://arxiv.org/abs/1908.08081}{arXiv:1908.08081 [hep-lat]}\BibitemShut
  {NoStop}%
\bibitem [{\citenamefont {Guo}\ and\ \citenamefont
  {D\"oring}(2020)}]{Guo:2019ogp}%
  \BibitemOpen
  \bibfield  {author} {\bibinfo {author} {\bibfnamefont {P.}~\bibnamefont
  {Guo}}\ and\ \bibinfo {author} {\bibfnamefont {M.}~\bibnamefont {D\"oring}},\
  }\bibfield  {title} {\enquote {\bibinfo {title} {{Lattice model of
  heavy-light three-body system}},}\ }\href {\doibase
  10.1103/PhysRevD.101.034501} {\bibfield  {journal} {\bibinfo  {journal}
  {Phys. Rev. D}\ }\textbf {\bibinfo {volume} {101}},\ \bibinfo {pages}
  {034501} (\bibinfo {year} {2020})},\ \Eprint
  {http://arxiv.org/abs/1910.08624}{arXiv:1910.08624 [hep-lat]}\BibitemShut
  {NoStop}%
\bibitem [{\citenamefont {Guo}(2020{\natexlab{b}})}]{Guo:2020wbl}%
  \BibitemOpen
  \bibfield  {author} {\bibinfo {author} {\bibfnamefont {P.}~\bibnamefont
  {Guo}},\ }\bibfield  {title} {\enquote {\bibinfo {title} {{Threshold
  expansion formula of $N$ bosons in a finite volume from a variational
  approach}},}\ }\href {\doibase 10.1103/PhysRevD.101.054512} {\bibfield
  {journal} {\bibinfo  {journal} {Phys. Rev. D}\ }\textbf {\bibinfo {volume}
  {101}},\ \bibinfo {pages} {054512} (\bibinfo {year} {2020}{\natexlab{b}})},\
  \Eprint {http://arxiv.org/abs/2002.04111}{arXiv:2002.04111
  [hep-lat]}\BibitemShut {NoStop}%
\bibitem [{\citenamefont {Guo}\ and\ \citenamefont
  {Long}(2020{\natexlab{a}})}]{Guo:2020kph}%
  \BibitemOpen
  \bibfield  {author} {\bibinfo {author} {\bibfnamefont {P.}~\bibnamefont
  {Guo}}\ and\ \bibinfo {author} {\bibfnamefont {B.}~\bibnamefont {Long}},\
  }\bibfield  {title} {\enquote {\bibinfo {title} {{Multi- $\pi^+$ systems in a
  finite volume}},}\ }\href {\doibase 10.1103/PhysRevD.101.094510} {\bibfield
  {journal} {\bibinfo  {journal} {Phys. Rev. D}\ }\textbf {\bibinfo {volume}
  {101}},\ \bibinfo {pages} {094510} (\bibinfo {year} {2020}{\natexlab{a}})},\
  \Eprint {http://arxiv.org/abs/2002.09266}{arXiv:2002.09266
  [hep-lat]}\BibitemShut {NoStop}%
\bibitem [{\citenamefont {Guo}(2020{\natexlab{c}})}]{Guo:2020iep}%
  \BibitemOpen
  \bibfield  {author} {\bibinfo {author} {\bibfnamefont {P.}~\bibnamefont
  {Guo}},\ }\href {https://arxiv.org/abs/2007.04473} {\enquote {\bibinfo
  {title} {Myth of scattering in finite volume},}\ } (\bibinfo {year}
  {2020}{\natexlab{c}}),\ \Eprint
  {http://arxiv.org/abs/2007.04473}{arXiv:2007.04473 [hep-lat]}\BibitemShut
  {NoStop}%
\bibitem [{\citenamefont {Guo}\ and\ \citenamefont
  {Long}(2020{\natexlab{b}})}]{Guo:2020ikh}%
  \BibitemOpen
  \bibfield  {author} {\bibinfo {author} {\bibfnamefont {P.}~\bibnamefont
  {Guo}}\ and\ \bibinfo {author} {\bibfnamefont {B.}~\bibnamefont {Long}},\
  }\bibfield  {title} {\enquote {\bibinfo {title} {{Visualizing resonances in
  finite volume}},}\ }\href {\doibase 10.1103/PhysRevD.102.074508} {\bibfield
  {journal} {\bibinfo  {journal} {Phys. Rev. D}\ }\textbf {\bibinfo {volume}
  {102}},\ \bibinfo {pages} {074508} (\bibinfo {year} {2020}{\natexlab{b}})},\
  \Eprint {http://arxiv.org/abs/2007.10895}{arXiv:2007.10895
  [hep-lat]}\BibitemShut {NoStop}%
\bibitem [{\citenamefont {Guo}(2020{\natexlab{d}})}]{Guo:2020spn}%
  \BibitemOpen
  \bibfield  {author} {\bibinfo {author} {\bibfnamefont {P.}~\bibnamefont
  {Guo}},\ }\bibfield  {title} {\enquote {\bibinfo {title} {{Modeling few-body
  resonances in finite volume}},}\ }\href {\doibase
  10.1103/PhysRevD.102.054514} {\bibfield  {journal} {\bibinfo  {journal}
  {Phys. Rev. D}\ }\textbf {\bibinfo {volume} {102}},\ \bibinfo {pages}
  {054514} (\bibinfo {year} {2020}{\natexlab{d}})},\ \Eprint
  {http://arxiv.org/abs/2007.12790}{arXiv:2007.12790 [hep-lat]}\BibitemShut
  {NoStop}%
\bibitem [{\citenamefont {Guo}\ and\ \citenamefont
  {Gasparian}(2021)}]{Guo:2021lhz}%
  \BibitemOpen
  \bibfield  {author} {\bibinfo {author} {\bibfnamefont {P.}~\bibnamefont
  {Guo}}\ and\ \bibinfo {author} {\bibfnamefont {V.}~\bibnamefont
  {Gasparian}},\ }\bibfield  {title} {\enquote {\bibinfo {title} {{Charged
  particles interaction in both a finite volume and a uniform magnetic
  field}},}\ }\href {\doibase 10.1103/PhysRevD.103.094520} {\bibfield
  {journal} {\bibinfo  {journal} {Phys. Rev. D}\ }\textbf {\bibinfo {volume}
  {103}},\ \bibinfo {pages} {094520} (\bibinfo {year} {2021})},\ \Eprint
  {http://arxiv.org/abs/2101.01150}{arXiv:2101.01150 [hep-lat]}\BibitemShut
  {NoStop}%
\bibitem [{\citenamefont {Guo}\ and\ \citenamefont {Long}(2022)}]{Guo:2021uig}%
  \BibitemOpen
  \bibfield  {author} {\bibinfo {author} {\bibfnamefont {P.}~\bibnamefont
  {Guo}}\ and\ \bibinfo {author} {\bibfnamefont {B.}~\bibnamefont {Long}},\
  }\bibfield  {title} {\enquote {\bibinfo {title} {{Nuclear reactions in
  artificial traps}},}\ }\href {\doibase 10.1088/1361-6471/ac59d5} {\bibfield
  {journal} {\bibinfo  {journal} {J. Phys. G}\ }\textbf {\bibinfo {volume}
  {49}},\ \bibinfo {pages} {055104} (\bibinfo {year} {2022})},\ \Eprint
  {http://arxiv.org/abs/2101.03901}{arXiv:2101.03901 [nucl-th]}\BibitemShut
  {NoStop}%
\bibitem [{\citenamefont {Guo}(2021)}]{Guo:2021qfu}%
  \BibitemOpen
  \bibfield  {author} {\bibinfo {author} {\bibfnamefont {P.}~\bibnamefont
  {Guo}},\ }\bibfield  {title} {\enquote {\bibinfo {title} {{Coulomb
  corrections to two-particle interactions in artificial traps}},}\ }\href
  {\doibase 10.1103/PhysRevC.103.064611} {\bibfield  {journal} {\bibinfo
  {journal} {Phys. Rev. C}\ }\textbf {\bibinfo {volume} {103}},\ \bibinfo
  {pages} {064611} (\bibinfo {year} {2021})},\ \Eprint
  {http://arxiv.org/abs/2101.11097}{arXiv:2101.11097 [nucl-th]}\BibitemShut
  {NoStop}%
\bibitem [{\citenamefont {Guo}\ and\ \citenamefont
  {Gasparian}(2022{\natexlab{a}})}]{Guo:2021hrf}%
  \BibitemOpen
  \bibfield  {author} {\bibinfo {author} {\bibfnamefont {P.}~\bibnamefont
  {Guo}}\ and\ \bibinfo {author} {\bibfnamefont {V.}~\bibnamefont
  {Gasparian}},\ }\bibfield  {title} {\enquote {\bibinfo {title} {{Charged
  particles interaction in both a finite volume and a uniform magnetic field
  II: topological and analytic properties of a magnetic system}},}\ }\href
  {\doibase 10.1088/1751-8121/ac7180} {\bibfield  {journal} {\bibinfo
  {journal} {J. Phys. A}\ }\textbf {\bibinfo {volume} {55}},\ \bibinfo {pages}
  {265201} (\bibinfo {year} {2022}{\natexlab{a}})},\ \Eprint
  {http://arxiv.org/abs/2107.10642}{arXiv:2107.10642 [hep-lat]}\BibitemShut
  {NoStop}%
\bibitem [{\citenamefont {Stetcu}\ \emph {et~al.}(2007)\citenamefont {Stetcu},
  \citenamefont {Barrett}, \citenamefont {van Kolck},\ and\ \citenamefont
  {Vary}}]{Stetcu:2007ms}%
  \BibitemOpen
  \bibfield  {author} {\bibinfo {author} {\bibfnamefont {I.}~\bibnamefont
  {Stetcu}}, \bibinfo {author} {\bibfnamefont {B.}~\bibnamefont {Barrett}},
  \bibinfo {author} {\bibfnamefont {U.}~\bibnamefont {van Kolck}}, \ and\
  \bibinfo {author} {\bibfnamefont {J.}~\bibnamefont {Vary}},\ }\bibfield
  {title} {\enquote {\bibinfo {title} {{Effective Theory for Trapped
  Few-Fermion Systems}},}\ }\href {\doibase 10.1103/PhysRevA.76.063613}
  {\bibfield  {journal} {\bibinfo  {journal} {Phys. Rev. A}\ }\textbf {\bibinfo
  {volume} {76}},\ \bibinfo {pages} {063613} (\bibinfo {year} {2007})},\
  \Eprint {http://arxiv.org/abs/0705.4335}{arXiv:0705.4335
  [cond-mat.other]}\BibitemShut {NoStop}%
\bibitem [{\citenamefont {Stetcu}\ \emph {et~al.}(2010)\citenamefont {Stetcu},
  \citenamefont {Rotureau}, \citenamefont {Barrett},\ and\ \citenamefont {van
  Kolck}}]{Stetcu:2010xq}%
  \BibitemOpen
  \bibfield  {author} {\bibinfo {author} {\bibfnamefont {I.}~\bibnamefont
  {Stetcu}}, \bibinfo {author} {\bibfnamefont {J.}~\bibnamefont {Rotureau}},
  \bibinfo {author} {\bibfnamefont {B.}~\bibnamefont {Barrett}}, \ and\
  \bibinfo {author} {\bibfnamefont {U.}~\bibnamefont {van Kolck}},\ }\bibfield
  {title} {\enquote {\bibinfo {title} {{An Effective field theory approach to
  two trapped particles}},}\ }\href {\doibase 10.1016/j.aop.2010.02.008}
  {\bibfield  {journal} {\bibinfo  {journal} {Annals Phys.}\ }\textbf {\bibinfo
  {volume} {325}},\ \bibinfo {pages} {1644} (\bibinfo {year} {2010})},\ \Eprint
  {http://arxiv.org/abs/1001.5071}{arXiv:1001.5071
  [cond-mat.quant-gas]}\BibitemShut {NoStop}%
\bibitem [{\citenamefont {Rotureau}\ \emph {et~al.}(2010)\citenamefont
  {Rotureau}, \citenamefont {Stetcu}, \citenamefont {Barrett}, \citenamefont
  {Birse},\ and\ \citenamefont {van Kolck}}]{Rotureau:2010uz}%
  \BibitemOpen
  \bibfield  {author} {\bibinfo {author} {\bibfnamefont {J.}~\bibnamefont
  {Rotureau}}, \bibinfo {author} {\bibfnamefont {I.}~\bibnamefont {Stetcu}},
  \bibinfo {author} {\bibfnamefont {B.}~\bibnamefont {Barrett}}, \bibinfo
  {author} {\bibfnamefont {M.}~\bibnamefont {Birse}}, \ and\ \bibinfo {author}
  {\bibfnamefont {U.}~\bibnamefont {van Kolck}},\ }\bibfield  {title} {\enquote
  {\bibinfo {title} {{Three and Four Harmonically Trapped Particles in an
  Effective Field Theory Framework}},}\ }\href {\doibase
  10.1103/PhysRevA.82.032711} {\bibfield  {journal} {\bibinfo  {journal} {Phys.
  Rev. A}\ }\textbf {\bibinfo {volume} {82}},\ \bibinfo {pages} {032711}
  (\bibinfo {year} {2010})},\ \Eprint
  {http://arxiv.org/abs/1006.3820}{arXiv:1006.3820
  [cond-mat.quant-gas]}\BibitemShut {NoStop}%
\bibitem [{\citenamefont {Rotureau}\ \emph {et~al.}(2012)\citenamefont
  {Rotureau}, \citenamefont {Stetcu}, \citenamefont {Barrett},\ and\
  \citenamefont {van Kolck}}]{Rotureau:2011vf}%
  \BibitemOpen
  \bibfield  {author} {\bibinfo {author} {\bibfnamefont {J.}~\bibnamefont
  {Rotureau}}, \bibinfo {author} {\bibfnamefont {I.}~\bibnamefont {Stetcu}},
  \bibinfo {author} {\bibfnamefont {B.}~\bibnamefont {Barrett}}, \ and\
  \bibinfo {author} {\bibfnamefont {U.}~\bibnamefont {van Kolck}},\ }\bibfield
  {title} {\enquote {\bibinfo {title} {{Two and Three Nucleons in a Trap and
  the Continuum Limit}},}\ }\href {\doibase 10.1103/PhysRevC.85.034003}
  {\bibfield  {journal} {\bibinfo  {journal} {Phys. Rev. C}\ }\textbf {\bibinfo
  {volume} {85}},\ \bibinfo {pages} {034003} (\bibinfo {year} {2012})},\
  \Eprint {http://arxiv.org/abs/1112.0267}{arXiv:1112.0267
  [nucl-th]}\BibitemShut {NoStop}%
\bibitem [{\citenamefont {Luu}\ \emph {et~al.}(2010)\citenamefont {Luu},
  \citenamefont {Savage}, \citenamefont {Schwenk},\ and\ \citenamefont
  {Vary}}]{Luu:2010hw}%
  \BibitemOpen
  \bibfield  {author} {\bibinfo {author} {\bibfnamefont {T.}~\bibnamefont
  {Luu}}, \bibinfo {author} {\bibfnamefont {M.~J.}\ \bibnamefont {Savage}},
  \bibinfo {author} {\bibfnamefont {A.}~\bibnamefont {Schwenk}}, \ and\
  \bibinfo {author} {\bibfnamefont {J.~P.}\ \bibnamefont {Vary}},\ }\bibfield
  {title} {\enquote {\bibinfo {title} {{Nucleon-Nucleon Scattering in a
  Harmonic Potential}},}\ }\href {\doibase 10.1103/PhysRevC.82.034003}
  {\bibfield  {journal} {\bibinfo  {journal} {Phys. Rev. C}\ }\textbf {\bibinfo
  {volume} {82}},\ \bibinfo {pages} {034003} (\bibinfo {year} {2010})},\
  \Eprint {http://arxiv.org/abs/1006.0427}{arXiv:1006.0427
  [nucl-th]}\BibitemShut {NoStop}%
\bibitem [{\citenamefont {Yang}(2016)}]{Yang:2016brl}%
  \BibitemOpen
  \bibfield  {author} {\bibinfo {author} {\bibfnamefont {C.-J.}\ \bibnamefont
  {Yang}},\ }\bibfield  {title} {\enquote {\bibinfo {title} {{Chiral potential
  renormalized in harmonic-oscillator space}},}\ }\href {\doibase
  10.1103/PhysRevC.94.064004} {\bibfield  {journal} {\bibinfo  {journal} {Phys.
  Rev. C}\ }\textbf {\bibinfo {volume} {94}},\ \bibinfo {pages} {064004}
  (\bibinfo {year} {2016})},\ \Eprint
  {http://arxiv.org/abs/1610.01350}{arXiv:1610.01350 [nucl-th]}\BibitemShut
  {NoStop}%
\bibitem [{\citenamefont {Johnson}\ \emph {et~al.}(2019)\citenamefont {Johnson}
  \emph {et~al.}}]{Johnson:2019sps}%
  \BibitemOpen
  \bibfield  {author} {\bibinfo {author} {\bibfnamefont {C.~W.}\ \bibnamefont
  {Johnson}} \emph {et~al.},\ }\bibfield  {title} {\enquote {\bibinfo {title}
  {{From bound states to the continuum}},}\ }in\ \href@noop {} {\emph {\bibinfo
  {booktitle} {{From Bound States to the Continuum}: {Connecting bound state
  calculations with scattering and reaction theory}}}}\ (\bibinfo {year}
  {2019})\ \Eprint {http://arxiv.org/abs/1912.00451}{arXiv:1912.00451
  [nucl-th]}\BibitemShut {NoStop}%
\bibitem [{\citenamefont {Zhang}(2020)}]{Zhang:2019cai}%
  \BibitemOpen
  \bibfield  {author} {\bibinfo {author} {\bibfnamefont {X.}~\bibnamefont
  {Zhang}},\ }\bibfield  {title} {\enquote {\bibinfo {title} {{Extracting
  free-space observables from trapped interacting clusters}},}\ }\href
  {\doibase 10.1103/PhysRevC.101.051602} {\bibfield  {journal} {\bibinfo
  {journal} {Phys. Rev. C}\ }\textbf {\bibinfo {volume} {101}},\ \bibinfo
  {pages} {051602} (\bibinfo {year} {2020})},\ \Eprint
  {http://arxiv.org/abs/1905.05275}{arXiv:1905.05275 [nucl-th]}\BibitemShut
  {NoStop}%
\bibitem [{\citenamefont {Zhang}\ \emph {et~al.}(2020)\citenamefont {Zhang},
  \citenamefont {Stroberg}, \citenamefont {Navr\'atil}, \citenamefont {Gwak},
  \citenamefont {Melendez}, \citenamefont {Furnstahl},\ and\ \citenamefont
  {Holt}}]{Zhang:2020rhz}%
  \BibitemOpen
  \bibfield  {author} {\bibinfo {author} {\bibfnamefont {X.}~\bibnamefont
  {Zhang}}, \bibinfo {author} {\bibfnamefont {S.}~\bibnamefont {Stroberg}},
  \bibinfo {author} {\bibfnamefont {P.}~\bibnamefont {Navr\'atil}}, \bibinfo
  {author} {\bibfnamefont {C.}~\bibnamefont {Gwak}}, \bibinfo {author}
  {\bibfnamefont {J.}~\bibnamefont {Melendez}}, \bibinfo {author}
  {\bibfnamefont {R.}~\bibnamefont {Furnstahl}}, \ and\ \bibinfo {author}
  {\bibfnamefont {J.}~\bibnamefont {Holt}},\ }\bibfield  {title} {\enquote
  {\bibinfo {title} {{Ab initio calculations of low-energy nuclear scattering
  using a generalized L\"uscher method}},}\ }\href {\doibase
  10.1103/PhysRevLett.125.112503} {\bibfield  {journal} {\bibinfo  {journal}
  {Phys. Rev. Lett.}\ }\textbf {\bibinfo {volume} {125}},\ \bibinfo {pages}
  {112503} (\bibinfo {year} {2020})},\ \Eprint
  {http://arxiv.org/abs/2004.13575}{arXiv:2004.13575 [nucl-th]}\BibitemShut
  {NoStop}%
\bibitem [{\citenamefont {Ishii}\ \emph {et~al.}(2007)\citenamefont {Ishii},
  \citenamefont {Aoki},\ and\ \citenamefont {Hatsuda}}]{PhysRevLett.99.022001}%
  \BibitemOpen
  \bibfield  {author} {\bibinfo {author} {\bibfnamefont {N.}~\bibnamefont
  {Ishii}}, \bibinfo {author} {\bibfnamefont {S.}~\bibnamefont {Aoki}}, \ and\
  \bibinfo {author} {\bibfnamefont {T.}~\bibnamefont {Hatsuda}},\ }\bibfield
  {title} {\enquote {\bibinfo {title} {Nuclear force from lattice qcd},}\
  }\href {\doibase 10.1103/PhysRevLett.99.022001} {\bibfield  {journal}
  {\bibinfo  {journal} {Phys. Rev. Lett.}\ }\textbf {\bibinfo {volume} {99}},\
  \bibinfo {pages} {022001} (\bibinfo {year} {2007})}\BibitemShut {NoStop}%
\bibitem [{\citenamefont {Aoki}\ \emph {et~al.}(2010)\citenamefont {Aoki},
  \citenamefont {Hatsuda},\ and\ \citenamefont {Ishii}}]{10.1143/PTP.123.89}%
  \BibitemOpen
  \bibfield  {author} {\bibinfo {author} {\bibfnamefont {S.}~\bibnamefont
  {Aoki}}, \bibinfo {author} {\bibfnamefont {T.}~\bibnamefont {Hatsuda}}, \
  and\ \bibinfo {author} {\bibfnamefont {N.}~\bibnamefont {Ishii}},\ }\bibfield
   {title} {\enquote {\bibinfo {title} {{Theoretical Foundation of the Nuclear
  Force in QCD and Its Applications to Central and Tensor Forces in Quenched
  Lattice QCD Simulations}},}\ }\href {\doibase 10.1143/PTP.123.89} {\bibfield
  {journal} {\bibinfo  {journal} {Progress of Theoretical Physics}\ }\textbf
  {\bibinfo {volume} {123}},\ \bibinfo {pages} {89} (\bibinfo {year} {2010})},\
  \Eprint
  {http://arxiv.org/abs/https://academic.oup.com/ptp/article-pdf/123/1/89/9681302/123-1-89.pdf}{https://academic.oup.com/ptp/article-pdf/123/1/89/9681302/123-1-89.pdf}\BibitemShut
  {NoStop}%
\bibitem [{\citenamefont {Iritani}\ \emph {et~al.}(2019)\citenamefont
  {Iritani}, \citenamefont {Aoki}, \citenamefont {Doi}, \citenamefont {Gongyo},
  \citenamefont {Hatsuda}, \citenamefont {Ikeda}, \citenamefont {Inoue},
  \citenamefont {Ishii}, \citenamefont {Nemura},\ and\ \citenamefont
  {Sasaki}}]{PhysRevD.99.014514}%
  \BibitemOpen
  \bibfield  {author} {\bibinfo {author} {\bibfnamefont {T.}~\bibnamefont
  {Iritani}}, \bibinfo {author} {\bibfnamefont {S.}~\bibnamefont {Aoki}},
  \bibinfo {author} {\bibfnamefont {T.}~\bibnamefont {Doi}}, \bibinfo {author}
  {\bibfnamefont {S.}~\bibnamefont {Gongyo}}, \bibinfo {author} {\bibfnamefont
  {T.}~\bibnamefont {Hatsuda}}, \bibinfo {author} {\bibfnamefont
  {Y.}~\bibnamefont {Ikeda}}, \bibinfo {author} {\bibfnamefont
  {T.}~\bibnamefont {Inoue}}, \bibinfo {author} {\bibfnamefont
  {N.}~\bibnamefont {Ishii}}, \bibinfo {author} {\bibfnamefont
  {H.}~\bibnamefont {Nemura}}, \ and\ \bibinfo {author} {\bibfnamefont
  {K.}~\bibnamefont {Sasaki}} (\bibinfo {collaboration} {HAL QCD
  Collaboration}),\ }\bibfield  {title} {\enquote {\bibinfo {title}
  {Systematics of the hal qcd potential at low energies in lattice qcd},}\
  }\href {\doibase 10.1103/PhysRevD.99.014514} {\bibfield  {journal} {\bibinfo
  {journal} {Phys. Rev. D}\ }\textbf {\bibinfo {volume} {99}},\ \bibinfo
  {pages} {014514} (\bibinfo {year} {2019})}\BibitemShut {NoStop}%
\bibitem [{\citenamefont {Ishii}\ \emph {et~al.}(2012)\citenamefont {Ishii},
  \citenamefont {Aoki}, \citenamefont {Doi}, \citenamefont {Hatsuda},
  \citenamefont {Ikeda}, \citenamefont {Inoue}, \citenamefont {Murano},
  \citenamefont {Nemura},\ and\ \citenamefont {Sasaki}}]{ISHII2012437}%
  \BibitemOpen
  \bibfield  {author} {\bibinfo {author} {\bibfnamefont {N.}~\bibnamefont
  {Ishii}}, \bibinfo {author} {\bibfnamefont {S.}~\bibnamefont {Aoki}},
  \bibinfo {author} {\bibfnamefont {T.}~\bibnamefont {Doi}}, \bibinfo {author}
  {\bibfnamefont {T.}~\bibnamefont {Hatsuda}}, \bibinfo {author} {\bibfnamefont
  {Y.}~\bibnamefont {Ikeda}}, \bibinfo {author} {\bibfnamefont
  {T.}~\bibnamefont {Inoue}}, \bibinfo {author} {\bibfnamefont
  {K.}~\bibnamefont {Murano}}, \bibinfo {author} {\bibfnamefont
  {H.}~\bibnamefont {Nemura}}, \ and\ \bibinfo {author} {\bibfnamefont
  {K.}~\bibnamefont {Sasaki}},\ }\bibfield  {title} {\enquote {\bibinfo {title}
  {Hadron–hadron interactions from imaginary-time nambu–bethe–salpeter
  wave function on the lattice},}\ }\href {\doibase
  https://doi.org/10.1016/j.physletb.2012.04.076} {\bibfield  {journal}
  {\bibinfo  {journal} {Physics Letters B}\ }\textbf {\bibinfo {volume}
  {712}},\ \bibinfo {pages} {437} (\bibinfo {year} {2012})}\BibitemShut
  {NoStop}%
\bibitem [{\citenamefont {Aoki}(2013)}]{AokiEPJA2013}%
  \BibitemOpen
  \bibfield  {author} {\bibinfo {author} {\bibfnamefont {S.}~\bibnamefont
  {Aoki}},\ }\bibfield  {title} {\enquote {\bibinfo {title} {Nucleon-nucleon
  interactions via lattice qcd: Methodology},}\ }\href {\doibase
  10.1140/epja/i2013-13081-0} {\bibfield  {journal} {\bibinfo  {journal} {The
  European Physical Journal A}\ }\textbf {\bibinfo {volume} {49}},\ \bibinfo
  {pages} {81} (\bibinfo {year} {2013})}\BibitemShut {NoStop}%
\bibitem [{\citenamefont {Lepage}(1989)}]{lepage1989analysis}%
  \BibitemOpen
  \bibfield  {author} {\bibinfo {author} {\bibfnamefont {G.~P.}\ \bibnamefont
  {Lepage}},\ }\bibfield  {title} {\enquote {\bibinfo {title} {The analysis of
  algorithms for lattice field theory},}\ }\href@noop {} {\bibfield  {journal}
  {\bibinfo  {journal} {Boulder ASI}\ }\textbf {\bibinfo {volume} {1989}},\
  \bibinfo {pages} {97} (\bibinfo {year} {1989})}\BibitemShut {NoStop}%
\bibitem [{\citenamefont {Drischler}\ \emph {et~al.}(2021)\citenamefont
  {Drischler}, \citenamefont {Haxton}, \citenamefont {McElvain}, \citenamefont
  {Mereghetti}, \citenamefont {Nicholson}, \citenamefont {Vranas},\ and\
  \citenamefont {Walker-Loud}}]{DRISCHLER2021103888}%
  \BibitemOpen
  \bibfield  {author} {\bibinfo {author} {\bibfnamefont {C.}~\bibnamefont
  {Drischler}}, \bibinfo {author} {\bibfnamefont {W.}~\bibnamefont {Haxton}},
  \bibinfo {author} {\bibfnamefont {K.}~\bibnamefont {McElvain}}, \bibinfo
  {author} {\bibfnamefont {E.}~\bibnamefont {Mereghetti}}, \bibinfo {author}
  {\bibfnamefont {A.}~\bibnamefont {Nicholson}}, \bibinfo {author}
  {\bibfnamefont {P.}~\bibnamefont {Vranas}}, \ and\ \bibinfo {author}
  {\bibfnamefont {A.}~\bibnamefont {Walker-Loud}},\ }\bibfield  {title}
  {\enquote {\bibinfo {title} {Towards grounding nuclear physics in qcd},}\
  }\href {\doibase https://doi.org/10.1016/j.ppnp.2021.103888} {\bibfield
  {journal} {\bibinfo  {journal} {Progress in Particle and Nuclear Physics}\
  }\textbf {\bibinfo {volume} {121}},\ \bibinfo {pages} {103888} (\bibinfo
  {year} {2021})}\BibitemShut {NoStop}%
\bibitem [{\citenamefont {Bulava}\ and\ \citenamefont
  {Hansen}(2019)}]{Bulava:2019kbi}%
  \BibitemOpen
  \bibfield  {author} {\bibinfo {author} {\bibfnamefont {J.}~\bibnamefont
  {Bulava}}\ and\ \bibinfo {author} {\bibfnamefont {M.~T.}\ \bibnamefont
  {Hansen}},\ }\bibfield  {title} {\enquote {\bibinfo {title} {{Scattering
  amplitudes from finite-volume spectral functions}},}\ }\href {\doibase
  10.1103/PhysRevD.100.034521} {\bibfield  {journal} {\bibinfo  {journal}
  {Phys. Rev.}\ }\textbf {\bibinfo {volume} {D100}},\ \bibinfo {pages} {034521}
  (\bibinfo {year} {2019})},\ \Eprint
  {http://arxiv.org/abs/1903.11735}{arXiv:1903.11735 [hep-lat]}\BibitemShut
  {NoStop}%
\bibitem [{\citenamefont {Guo}\ \emph {et~al.}(2024)\citenamefont {Guo},
  \citenamefont {Gasparian}, \citenamefont {P\'erez-Garrido},\ and\
  \citenamefont {J\'odar}}]{Guo:2024bar}%
  \BibitemOpen
  \bibfield  {author} {\bibinfo {author} {\bibfnamefont {P.}~\bibnamefont
  {Guo}}, \bibinfo {author} {\bibfnamefont {V.}~\bibnamefont {Gasparian}},
  \bibinfo {author} {\bibfnamefont {A.}~\bibnamefont {P\'erez-Garrido}}, \ and\
  \bibinfo {author} {\bibfnamefont {E.}~\bibnamefont {J\'odar}},\ }\bibfield
  {title} {\enquote {\bibinfo {title} {{Tunneling time in coupled-channel
  systems}},}\ }\href {\doibase 10.1103/PhysRevResearch.6.043032} {\bibfield
  {journal} {\bibinfo  {journal} {Phys. Rev. Res.}\ }\textbf {\bibinfo {volume}
  {6}},\ \bibinfo {pages} {043032} (\bibinfo {year} {2024})},\ \Eprint
  {http://arxiv.org/abs/2407.17981}{arXiv:2407.17981
  [cond-mat.other]}\BibitemShut {NoStop}%
\bibitem [{\citenamefont {Guo}(2025)}]{Guo:2025vgk}%
  \BibitemOpen
  \bibfield  {author} {\bibinfo {author} {\bibfnamefont {P.}~\bibnamefont
  {Guo}},\ }\href {https://arxiv.org/abs/2504.14474} {\enquote {\bibinfo
  {title} {Toward extracting scattering phase shift from integrated correlation
  functions on quantum computers},}\ } (\bibinfo {year} {2025}),\ \Eprint
  {http://arxiv.org/abs/2504.14474}{arXiv:2504.14474 [quant-ph]}\BibitemShut
  {NoStop}%
\bibitem [{\citenamefont {Zhang}\ \emph
  {et~al.}(2024{\natexlab{a}})\citenamefont {Zhang}, \citenamefont {Bai},
  \citenamefont {Wang},\ and\ \citenamefont {Ren}}]{Zhang:2024mot}%
  \BibitemOpen
  \bibfield  {author} {\bibinfo {author} {\bibfnamefont {H.}~\bibnamefont
  {Zhang}}, \bibinfo {author} {\bibfnamefont {D.}~\bibnamefont {Bai}}, \bibinfo
  {author} {\bibfnamefont {Z.}~\bibnamefont {Wang}}, \ and\ \bibinfo {author}
  {\bibfnamefont {Z.}~\bibnamefont {Ren}},\ }\bibfield  {title} {\enquote
  {\bibinfo {title} {{Charged particle scattering in harmonic traps}},}\ }\href
  {\doibase 10.1016/j.physletb.2024.138490} {\bibfield  {journal} {\bibinfo
  {journal} {Phys. Lett. B}\ }\textbf {\bibinfo {volume} {850}},\ \bibinfo
  {pages} {138490} (\bibinfo {year} {2024}{\natexlab{a}})}\BibitemShut
  {NoStop}%
\bibitem [{\citenamefont {Bagnarol}\ \emph {et~al.}(2025)\citenamefont
  {Bagnarol}, \citenamefont {Barnea}, \citenamefont {Rojik},\ and\
  \citenamefont {Schafer}}]{Bagnarol:2024rhq}%
  \BibitemOpen
  \bibfield  {author} {\bibinfo {author} {\bibfnamefont {M.}~\bibnamefont
  {Bagnarol}}, \bibinfo {author} {\bibfnamefont {N.}~\bibnamefont {Barnea}},
  \bibinfo {author} {\bibfnamefont {M.}~\bibnamefont {Rojik}}, \ and\ \bibinfo
  {author} {\bibfnamefont {M.}~\bibnamefont {Schafer}},\ }\bibfield  {title}
  {\enquote {\bibinfo {title} {{Accurate calculation of low energy scattering
  phase shifts of charged particles in a harmonic oscillator trap}},}\ }\href
  {\doibase 10.1016/j.physletb.2024.139230} {\bibfield  {journal} {\bibinfo
  {journal} {Phys. Lett. B}\ }\textbf {\bibinfo {volume} {861}},\ \bibinfo
  {pages} {139230} (\bibinfo {year} {2025})},\ \Eprint
  {http://arxiv.org/abs/2410.02602}{arXiv:2410.02602 [nucl-th]}\BibitemShut
  {NoStop}%
\bibitem [{\citenamefont {Zhang}\ \emph
  {et~al.}(2024{\natexlab{b}})\citenamefont {Zhang}, \citenamefont {Bai},\ and\
  \citenamefont {Ren}}]{Zhang:2024vch}%
  \BibitemOpen
  \bibfield  {author} {\bibinfo {author} {\bibfnamefont {H.}~\bibnamefont
  {Zhang}}, \bibinfo {author} {\bibfnamefont {D.}~\bibnamefont {Bai}}, \ and\
  \bibinfo {author} {\bibfnamefont {Z.}~\bibnamefont {Ren}},\ }\bibfield
  {title} {\enquote {\bibinfo {title} {{Coupled-channels reactions for charged
  particles in harmonic traps}},}\ }\href {\doibase
  10.1103/PhysRevC.110.034308} {\bibfield  {journal} {\bibinfo  {journal}
  {Phys. Rev. C}\ }\textbf {\bibinfo {volume} {110}},\ \bibinfo {pages}
  {034308} (\bibinfo {year} {2024}{\natexlab{b}})}\BibitemShut {NoStop}%
\bibitem [{\citenamefont {Beane}\ and\ \citenamefont
  {Savage}(2014)}]{Beane_2014}%
  \BibitemOpen
  \bibfield  {author} {\bibinfo {author} {\bibfnamefont {S.~R.}\ \bibnamefont
  {Beane}}\ and\ \bibinfo {author} {\bibfnamefont {M.~J.}\ \bibnamefont
  {Savage}},\ }\bibfield  {title} {\enquote {\bibinfo {title} {Two-particle
  elastic scattering in a finite volume including qed},}\ }\href {\doibase
  10.1103/physrevd.90.074511} {\bibfield  {journal} {\bibinfo  {journal}
  {Physical Review D}\ }\textbf {\bibinfo {volume} {90}} (\bibinfo {year}
  {2014}),\ 10.1103/physrevd.90.074511}\BibitemShut {NoStop}%
\bibitem [{\citenamefont {Beane}\ \emph {et~al.}(2021)\citenamefont {Beane}
  \emph {et~al.}}]{NPLQCD:2020ozd}%
  \BibitemOpen
  \bibfield  {author} {\bibinfo {author} {\bibfnamefont {S.~R.}\ \bibnamefont
  {Beane}} \emph {et~al.} (\bibinfo {collaboration} {NPLQCD, QCDSF}),\
  }\bibfield  {title} {\enquote {\bibinfo {title} {{Charged multihadron systems
  in lattice QCD+QED}},}\ }\href {\doibase 10.1103/PhysRevD.103.054504}
  {\bibfield  {journal} {\bibinfo  {journal} {Phys. Rev. D}\ }\textbf {\bibinfo
  {volume} {103}},\ \bibinfo {pages} {054504} (\bibinfo {year} {2021})},\
  \Eprint {http://arxiv.org/abs/2003.12130}{arXiv:2003.12130
  [hep-lat]}\BibitemShut {NoStop}%
\bibitem [{\citenamefont {Yu}\ \emph {et~al.}(2023)\citenamefont {Yu},
  \citenamefont {K\"onig},\ and\ \citenamefont {Lee}}]{Yu:2022nzm}%
  \BibitemOpen
  \bibfield  {author} {\bibinfo {author} {\bibfnamefont {H.}~\bibnamefont
  {Yu}}, \bibinfo {author} {\bibfnamefont {S.}~\bibnamefont {K\"onig}}, \ and\
  \bibinfo {author} {\bibfnamefont {D.}~\bibnamefont {Lee}},\ }\bibfield
  {title} {\enquote {\bibinfo {title} {{Charged-Particle Bound States in
  Periodic Boxes}},}\ }\href {\doibase 10.1103/PhysRevLett.131.212502}
  {\bibfield  {journal} {\bibinfo  {journal} {Phys. Rev. Lett.}\ }\textbf
  {\bibinfo {volume} {131}},\ \bibinfo {pages} {212502} (\bibinfo {year}
  {2023})},\ \Eprint {http://arxiv.org/abs/2212.14379}{arXiv:2212.14379
  [nucl-th]}\BibitemShut {NoStop}%
\bibitem [{\citenamefont {Bubna}\ \emph {et~al.}(2024)\citenamefont {Bubna},
  \citenamefont {Hammer}, \citenamefont {Müller}, \citenamefont {Pang},
  \citenamefont {Rusetsky},\ and\ \citenamefont {Wu}}]{bubna2024}%
  \BibitemOpen
  \bibfield  {author} {\bibinfo {author} {\bibfnamefont {R.}~\bibnamefont
  {Bubna}}, \bibinfo {author} {\bibfnamefont {H.-W.}\ \bibnamefont {Hammer}},
  \bibinfo {author} {\bibfnamefont {F.}~\bibnamefont {Müller}}, \bibinfo
  {author} {\bibfnamefont {J.-Y.}\ \bibnamefont {Pang}}, \bibinfo {author}
  {\bibfnamefont {A.}~\bibnamefont {Rusetsky}}, \ and\ \bibinfo {author}
  {\bibfnamefont {J.-J.}\ \bibnamefont {Wu}},\ }\href
  {https://arxiv.org/abs/2402.12985} {\enquote {\bibinfo {title} {L\"uscher
  equation with long-range forces},}\ } (\bibinfo {year} {2024}),\ \Eprint
  {http://arxiv.org/abs/2402.12985}{arXiv:2402.12985 [hep-lat]}\BibitemShut
  {NoStop}%
\bibitem [{\citenamefont {Cornwell}(1997)}]{Cornwell:1997ke}%
  \BibitemOpen
  \bibfield  {author} {\bibinfo {author} {\bibfnamefont {J.~F.}\ \bibnamefont
  {Cornwell}},\ }\href@noop {} {\emph {\bibinfo {title} {{Group theory in
  physics: An introduction}}}}\ (\bibinfo  {publisher} {San Diego, California,
  USA: Academic Press},\ \bibinfo {year} {1997})\BibitemShut {NoStop}%
\bibitem [{\citenamefont {Poliatzky}(1993)}]{Poliatzky:1992gn}%
  \BibitemOpen
  \bibfield  {author} {\bibinfo {author} {\bibfnamefont {N.}~\bibnamefont
  {Poliatzky}},\ }\bibfield  {title} {\enquote {\bibinfo {title}
  {{Normalization of scattering states, scattering phase shifts and Levinson's
  theorem}},}\ }\href@noop {} {\bibfield  {journal} {\bibinfo  {journal} {Helv.
  Phys. Acta}\ }\textbf {\bibinfo {volume} {66}},\ \bibinfo {pages} {241}
  (\bibinfo {year} {1993})},\ \Eprint
  {http://arxiv.org/abs/hep-th/9304008}{arXiv:hep-th/9304008}\BibitemShut
  {NoStop}%
\bibitem [{\citenamefont {Guo}\ and\ \citenamefont
  {Gasparian}(2022{\natexlab{b}})}]{Guo:2022row}%
  \BibitemOpen
  \bibfield  {author} {\bibinfo {author} {\bibfnamefont {P.}~\bibnamefont
  {Guo}}\ and\ \bibinfo {author} {\bibfnamefont {V.}~\bibnamefont
  {Gasparian}},\ }\bibfield  {title} {\enquote {\bibinfo {title} {{Friedel
  formula and Krein's theorem in complex potential scattering theory}},}\
  }\href {\doibase 10.1103/PhysRevResearch.4.023083} {\bibfield  {journal}
  {\bibinfo  {journal} {Phys. Rev. Res.}\ }\textbf {\bibinfo {volume} {4}},\
  \bibinfo {pages} {023083} (\bibinfo {year} {2022}{\natexlab{b}})},\ \Eprint
  {http://arxiv.org/abs/2202.12465}{arXiv:2202.12465
  [cond-mat.other]}\BibitemShut {NoStop}%
\bibitem [{\citenamefont {Friedel}(1958)}]{Friedel1958}%
  \BibitemOpen
  \bibfield  {author} {\bibinfo {author} {\bibfnamefont {J.}~\bibnamefont
  {Friedel}},\ }\bibfield  {title} {\enquote {\bibinfo {title} {Metallic
  alloys},}\ }\href {\doibase 10.1007/BF02751483} {\bibfield  {journal}
  {\bibinfo  {journal} {Il Nuovo Cimento (1955-1965)}\ }\textbf {\bibinfo
  {volume} {7}},\ \bibinfo {pages} {287} (\bibinfo {year} {1958})}\BibitemShut
  {NoStop}%
\bibitem [{\citenamefont {{Birman}}\ and\ \citenamefont
  {{Kre\u{\i}n}}(1962)}]{zbMATH03313022}%
  \BibitemOpen
  \bibfield  {author} {\bibinfo {author} {\bibfnamefont {M.~S.}\ \bibnamefont
  {{Birman}}}\ and\ \bibinfo {author} {\bibfnamefont {M.~G.}\ \bibnamefont
  {{Kre\u{\i}n}}},\ }\bibfield  {title} {{\selectlanguage {English}\enquote
  {\bibinfo {title} {{On the theory of wave operators and scattering
  operators}},}\ }}\href@noop {} {\bibfield  {journal} {\bibinfo  {journal}
  {{Sov. Math., Dokl.}}\ }\textbf {\bibinfo {volume} {3}},\ \bibinfo {pages}
  {740} (\bibinfo {year} {1962})}\BibitemShut {NoStop}%
\bibitem [{\citenamefont {Krein}(1953)}]{krein1953trace}%
  \BibitemOpen
  \bibfield  {author} {\bibinfo {author} {\bibfnamefont {M.~G.}\ \bibnamefont
  {Krein}},\ }\bibfield  {title} {\enquote {\bibinfo {title} {On the trace
  formula in perturbation theory},}\ }\href@noop {} {\bibfield  {journal}
  {\bibinfo  {journal} {Matematicheskii sbornik}\ }\textbf {\bibinfo {volume}
  {75}},\ \bibinfo {pages} {597} (\bibinfo {year} {1953})}\BibitemShut
  {NoStop}%
\bibitem [{\citenamefont {Messiah}(1999)}]{messiah1999quantum}%
  \BibitemOpen
  \bibfield  {author} {\bibinfo {author} {\bibfnamefont {A.}~\bibnamefont
  {Messiah}},\ }\href {https://books.google.com/books?id=mwssSDXzkNcC} {\emph
  {\bibinfo {title} {Quantum Mechanics}}},\ Dover books on physics\ (\bibinfo
  {publisher} {Dover Publications},\ \bibinfo {year} {1999})\BibitemShut
  {NoStop}%
\bibitem [{\citenamefont {Rothkopf}(2017)}]{Rothkopf_2017}%
  \BibitemOpen
  \bibfield  {author} {\bibinfo {author} {\bibfnamefont {A.}~\bibnamefont
  {Rothkopf}},\ }\bibfield  {title} {\enquote {\bibinfo {title} {Bayesian
  inference of nonpositive spectral functions in quantum field theory},}\
  }\href {\doibase 10.1103/physrevd.95.056016} {\bibfield  {journal} {\bibinfo
  {journal} {Physical Review D}\ }\textbf {\bibinfo {volume} {95}} (\bibinfo
  {year} {2017}),\ 10.1103/physrevd.95.056016}\BibitemShut {NoStop}%
\bibitem [{\citenamefont {Rothkopf}(2022)}]{Rothkopf_2022}%
  \BibitemOpen
  \bibfield  {author} {\bibinfo {author} {\bibfnamefont {A.}~\bibnamefont
  {Rothkopf}},\ }\bibfield  {title} {\enquote {\bibinfo {title} {Bayesian
  inference of real-time dynamics from lattice qcd},}\ }\href {\doibase
  10.3389/fphy.2022.1028995} {\bibfield  {journal} {\bibinfo  {journal}
  {Frontiers in Physics}\ }\textbf {\bibinfo {volume} {10}} (\bibinfo {year}
  {2022}),\ 10.3389/fphy.2022.1028995}\BibitemShut {NoStop}%
\bibitem [{\citenamefont {Yang}\ \emph {et~al.}(2023)\citenamefont {Yang},
  \citenamefont {Du},\ and\ \citenamefont {Huang}}]{yang2023}%
  \BibitemOpen
  \bibfield  {author} {\bibinfo {author} {\bibfnamefont {S.}~\bibnamefont
  {Yang}}, \bibinfo {author} {\bibfnamefont {L.}~\bibnamefont {Du}}, \ and\
  \bibinfo {author} {\bibfnamefont {L.}~\bibnamefont {Huang}},\ }\href
  {https://arxiv.org/abs/2401.00018} {\enquote {\bibinfo {title} {Combining
  bayesian reconstruction entropy with maximum entropy method for analytic
  continuations of matrix-valued green's functions},}\ } (\bibinfo {year}
  {2023}),\ \Eprint {http://arxiv.org/abs/2401.00018}{arXiv:2401.00018
  [hep-lat]}\BibitemShut {NoStop}%
\bibitem [{\citenamefont {Hostler}(1964)}]{Hostler1964}%
  \BibitemOpen
  \bibfield  {author} {\bibinfo {author} {\bibfnamefont {L.}~\bibnamefont
  {Hostler}},\ }\bibfield  {title} {\enquote {\bibinfo {title} {Coulomb green's
  functions and the furry approximation},}\ }\href {\doibase 10.1063/1.1704153}
  {\bibfield  {journal} {\bibinfo  {journal} {Journal of Mathematical Physics}\
  }\textbf {\bibinfo {volume} {5}},\ \bibinfo {pages} {591} (\bibinfo {year}
  {1964})},\ \Eprint
  {http://arxiv.org/abs/https://doi.org/10.1063/1.1704153}{https://doi.org/10.1063/1.1704153}\BibitemShut
  {NoStop}%
\end{thebibliography}%

\end{document}